\documentclass[twocolumn,twocolappendix]{aastex7}

\usepackage{bm}
\usepackage{mathtools}

\usepackage{color}

\begin{document}
\title{The \emph{R}-Process Alliance: Enrichment of \emph{r}-process Elements in a Simulated Milky Way-like Galaxy}
\shorttitle{Galactic \emph{r}-process enrichment}
\shortauthors{Hirai et al.}
\correspondingauthor{Yutaka Hirai}

\author[0000-0002-5661-033X]{Yutaka Hirai}
\affiliation{Department of Community Service and Science, Tohoku University of Community Service and Science, 3-5-1 Iimoriyama, Sakata, Yamagata 998-8580, Japan}
\email[show]{yutaka.hirai@koeki-u.ac.jp}

\author[0000-0003-4573-6233]{Timothy C.\ Beers}
\affiliation{Department of Physics and Astronomy, University of Notre Dame,
225 Nieuwland Science Hall, Notre Dame, IN 46556, USA}
\affiliation{Joint Institute for Nuclear Astrophysics, Center for the Evolution of the Elements (JINA-CEE), East Lansing, MI 48824, USA}
\email{tbeers@nd.edu}

\author[0000-0001-5297-4518]{Young Sun Lee}
\affiliation{Department of Astronomy and Space Science, Chungnam National University, Daejeon 34134, Republic of Korea}
\email{youngsun@cnu.ac.kr}

\author[0000-0002-4759-7794]{Shinya Wanajo}
\affiliation{Astronomical Institute, Tohoku University, 6-3 Aoba, Aramaki, Aoba-ku, Sendai, Miyagi 980-8578, Japan}
\affiliation{RIKEN iTHEMS Research Group, 2-1 Hirosawa, Wako, Saitama 351-0198, Japan}
\email{shinya.wanajo@astr.tohoku.ac.jp}

\author[0000-0001-5107-8930]{Ian U.\ Roederer}
\affiliation{Department of Physics, North Carolina State University,
2401 Stinson Dr, Box 8202, Raleigh, NC 27695, USA}
\affiliation{Joint Institute for Nuclear Astrophysics, Center for the Evolution of the Elements (JINA-CEE), East Lansing, MI 48824, USA}
\email{iuroederer@ncsu.edu}

\author[0000-0001-8253-6850]{Masaomi Tanaka}
\affiliation{Astronomical Institute, Tohoku University, 6-3 Aoba, Aramaki, Aoba-ku, Sendai, Miyagi 980-8578, Japan}
\affiliation{Division for the Establishment of Frontier Sciences, Organization for Advanced Studies, Tohoku University, Sendai, Miyagi 980-8577, Japan}
\email{masaomi.tanaka@astr.tohoku.ac.jp}

\author[0000-0002-9053-860X]{Masashi Chiba}
\affiliation{Astronomical Institute, Tohoku University, 6-3 Aoba, Aramaki, Aoba-ku, Sendai, Miyagi 980-8578, Japan}
\email{chiba@astr.tohoku.ac.jp}

\author[0000-0001-8226-4592]{Takayuki R.\ Saitoh}
\affiliation{Department of Planetology, Graduate School of Science, Kobe University, 1-1 Rokkodai-cho, Nada-ku, Kobe, Hyogo 657-8501, Japan}
\affiliation{Center for Planetary Science (CPS), Graduate School of Science, Kobe University 1-1 Rokkodai, Nada-ku, Kobe, Hyogo 657-8501, Japan}
\email{saitoh@people.kobe-u.ac.jp}

\author[0000-0003-4479-1265]{Vinicius M.\ Placco}
\affiliation{NSF NOIRLab, Tucson, AZ 85719, USA}
\email{vinicius.placco@noirlab.edu}

\author[0000-0001-6154-8983]{Terese T.\ Hansen}
\affiliation{Department of Astronomy, Stockholm University, AlbaNova University Center, SE-106 91 Stockholm, Sweden}
\email{thidemannhansen@gmail.com}

\author[0000-0002-8504-8470]{Rana Ezzeddine}
\affiliation{Department of Astronomy, University of Florida, Bryant Space Science Center, Gainesville, FL 32611, USA}
\affiliation{Joint Institute for Nuclear Astrophysics, Center for the Evolution of the Elements (JINA-CEE), East Lansing, MI 48824, USA}
\email{rezzeddine@ufl.edu}

\author[0000-0002-2139-7145]{Anna Frebel}
\affiliation{Department of Physics and Kavli Institute for Astrophysics and Space Research, Massachusetts Institute of Technology, Cambridge, MA 02139, USA}
\affiliation{Joint Institute for Nuclear Astrophysics, Center for the Evolution of the Elements (JINA-CEE), East Lansing, MI 48824, USA}
\email{afrebel@mit.edu}

\author[0000-0002-5463-6800]{Erika M.\ Holmbeck}
\affiliation{Lawrence Livermore National Laboratory, 7000 East Avenue, Livermore, CA 94550, USA}
\affiliation{Joint Institute for Nuclear Astrophysics, Center for the Evolution of the Elements (JINA-CEE), East Lansing, MI 48824, USA}
\email{holmbeck1@llnl.gov}

\author[0000-0002-5095-4000]{Charli M.\ Sakari}
\affiliation{Department of Physics and Astronomy, San Francisco State University, San Francisco, CA 94132, USA}
\email{sakaricm@sfsu.edu}

\begin{abstract}
We study the formation of stars with varying amounts of heavy elements synthesized by the rapid neutron-capture process (\emph{r}-process) based on our detailed cosmological zoom-in simulation of a Milky Way-like galaxy with an $N$-body/smoothed particle hydrodynamics code, \textsc{asura}.
Most stars with no overabundance in \emph{r}-process elements, as well as the strongly \emph{r}-process enhanced \emph{r}-II stars ([Eu/Fe] $>+0.7$), are formed in dwarf galaxies accreted by the Milky Way within the 6 Gyr after the Big Bang. In contrast, over half of the moderately enhanced \emph{r}-I stars ($+0.3 <$ [Eu/Fe] $\leq +0.7$) are formed in the main \textit{in-situ} disk after 6 Gyr. 
Our results suggest that the fraction of \emph{r}-I and \emph{r}-II stars formed in disrupted dwarf galaxies is larger the higher their [Eu/Fe] is. Accordingly, the most strongly enhanced \emph{r}-III stars ([Eu/Fe] $> +2.0$) are formed in accreted components. 
These results suggest that non-\emph{r}-process-enhanced stars and \emph{r}-II stars are mainly formed in low-mass dwarf galaxies that hosted either none or a single neutron star merger, while the \emph{r}-I stars tend to form in the well-mixed \textit{in-situ} disk. 
We compare our findings with high-resolution spectroscopic observations of \emph{r}-process-enhanced metal-poor stars in the halo and dwarf galaxies, including those collected by the R-Process Alliance.
We conclude that observed [Eu/Fe] and [Eu/Mg] ratios can be employed in chemical tagging of the Milky Way's accretion history.
\end{abstract}

\section{Introduction} \label{sec:intro}

The abundances of the elements synthesized by the rapid neutron-capture process 
(\emph{r}-process) are part of the fossil record of the early Universe preserved in the most metal-poor stars. The Milky Way (MW)'s very metal-poor (VMP) stars with [Fe/H]\footnote{[X/Y] = log($N_{\rm{x}}$/$N_{\rm{y}}$)$-$log($N_{\rm{x}}$/$N_{\rm{y}}$)$_{\sun}$, where $N_{\rm{x}}$ and $N_{\rm{y}}$ represent the number density of the elements X and Y.}
$<-$2 exhibit a large range in their \emph{r}-process-to-Fe ratios \citep[e.g.,][]{Honda2004, Holmbeck2020}. The abundance patterns of the lanthanides ($56 \leq Z \leq 71$) and the third-peak ($76 \leq Z \leq 88$) \emph{r}-process elements of stars with [Eu/Fe] $>+$0.7 (\emph{r}-II stars) closely match that of the scaled Solar System \emph{r}-process abundances \citep[e.g.,][]{Sneden2003, Frebel2007, Roederer2022}. This apparent universality shared by \emph{r}-process stars provides the opportunity to study the origins of these \emph{r}-process element enrichment and to use these stars as tracers of the formation process of the MW. 

Ongoing high-resolution spectroscopic observations have greatly increased the known metal-poor stars with derived abundance patterns of \emph{r}-process elements. The \emph{R}-Process Alliance (RPA), in particular, has reported on chemical abundance analyses of 595 metal-poor stars in their five data releases \citep{Hansen2018, Sakari2018, Ezzeddine2020, Holmbeck2020, Bandyopadhyay2024}. Supplemented by astrometric data from \textit{Gaia} {\citep{Gaia2016}}, the RPA has also identified numerous chemo-dynamically tagged groups (CDTGs) of  \emph{r}-process enhanced (RPE) stars \citep{Gudin2021, Shank2023} within the halo and disk populations of the MW, which may be remnants of disrupted dwarf galaxies.

The \emph{r}-process abundances in metal-poor stars are directly linked to the ejecta of the astrophysical site(s) of the \emph{r}-process. \citet{Roederer2023} argued that the \emph{r}-process elements with mass numbers $A$ = 99 to 110 correlate with heavier ones ($A>$150), indicating that the fission of transuranic nuclei ($A>$ 260) occurs in \emph{r}-process events. Binary neutron star mergers (NSMs) are likely sites of such nucleosynthetic environments \citep[e.g.,][]{Wanajo14, Radice2016, Fujibayashi2023}. Observations of the electromagnetic counterpart {(kilonova)} of the NSM GW170817{, along with the radiative-transfer simulations of kilonovae,} support the existence of lanthanides in its ejecta \citep[e.g.,][]{Abbott2017PhRvL,Abbott2017ApJ,Arcavi2017, Drout2017, Nicholl2017, Shappee2017, Tanaka2017}. {Spectroscopic analyses of this kilonova also suggest the presence of several neutron-capture elements in the ejecta, such as Sr \citep{Watson2019}, Y \citep{Sneppen2023}, Te \citep{Hotokezaka2023}, La, and Ce \citep{Domoto2022}.} Magnetorotational core-collapse supernovae \citep[CCSNe, e.g.,][]{Winteler2012, Nishimura2015, yong21} and collapsars \citep{Siegel2019}
 may also synthesize \emph{r}-process elements.

The enrichment of \emph{r}-process elements in galaxies can be understood by comparing chemical-evolution models and observations. Most previous studies focused on understanding the astrophysical sites of the \emph{r}-process \citep[e.g.,][]{Mathews1990, Argast2004,  Cescutti2014, Matteucci2014, Haynes2019, Cote2019, 
vandeVoort2020, Cavallo2023}. \citet{Argast2004} described the difficulty of reproducing the observed [Eu/Fe] ratios with yields from NSMs due to their low rates and long delay times. This problem can be resolved by taking into account the enrichment of \emph{r}-process elements in the MW's satellite dwarf galaxies, which serve as probes of galaxy assembly processes \citep[e.g.,][]{Ishimaru2015, Shen2015, vandeVoort2015, Hirai2015, Hirai2017,   Ojima2018, Wanajo2021}. 

As one example, the small \emph{r}-process enriched ultra-faint dwarf (UFD) galaxy Reticulum\,II \citep{Ji2016, Roederer2016} is an ideal probe because Reticulum\,II was likely enriched by an NSM at the earliest times. The observations of seven RPE stars in this system presented the first observational evidence in support of a single \emph{r}-process source, such as an NSM. Additional RPE stars in Reticulum\,II have recently been reported by \cite{Ji2023}.
 
According to hierarchical structure formation models, numerical simulations, and observational evidence, the MW comprises stars formed in the main halo of the MW (\textit{in-situ} component) and in satellite galaxies accreted later \citep[accreted component, e.g.,][]{Carollo2007,Carollo2010,Beers2012,Tissera2014,Font2020}. Identifying \textit{where} RPE stars currently present in the halo were formed provides important information about the MW's assembly history \citep[e.g.,][]{Brauer2019, Hirai2022}. Early work by \citet{Roederer2018} suggested that some RPE stars exhibit common kinematics and, hence, origins. \citet{Hirai2022} showed that most \emph{r}-II stars with [Fe/H] $< -$2.5 come from accreted components based on their cosmological zoom-in simulation.  With the discovery of the many more known RPE stars in recent years, refined statistical analyses by \citet{Gudin2021} and \citet{Shank2023} demonstrated that the RPE stars have likely share common chemical-evolution histories, presumably in their parent satellite galaxies or in globular clusters, long before they were disrupted into the MW’s halo \citep[see also][]{Aguado2021,Matsuno2021, Gull2021, Naidu2022, Monty2024}. \citet{Hattori2023} found six groups of \emph{r}-II stars with similar orbits and chemical abundances, suggesting that they come from accreted components.

In order to reconstruct the MW's assembly history using \emph{r}-process abundances as a probe, we need to clarify where and how stars with \emph{r}-process elements are formed during their evolutionary history. 
Following \citet{Beers2005}, \citet{Holmbeck2020}, and references therein, stars are classified with [Eu/Fe] ratios into three broad categories, based on their level of \emph{r}-process enhancement: non \emph{r}-process-enhanced (non-RPE; [Eu/Fe] $\leq+0.3$), moderately RPE (\emph{r}-I; $+0.3 <$ [Eu/Fe] $\leq +0.7$), and highly RPE (\emph{r}-II; [Eu/Fe] $>+0.7$). 
Recently, the \emph{r}-III category of \emph{r}-process enhancement has been suggested (\emph{r}-III; [Eu/Fe] $> +2.0$; \citealt{Cain2020}), but only a handful of such stars are presently known (e.g., \citealt{Cain2020,Roederer2024}).

This study aims to understand the formation environments of non-RPE, \emph{r}-I, and \emph{r}-II stars using a cosmological zoom-in simulation of a MW-like galaxy with an $N$-body/smoothed particle hydrodynamics (SPH) code, \textsc{asura}. We explore these stars' formation times and birthplaces across the early universe and characterize the distribution of \emph{r}-process elements in \textit{in-situ} and accreted components of the emerging Milky Way.

This paper is organized as follows. Section \ref{sec:method} describes our code and the initial conditions adopted in this study. In Section \ref{sec:results}, we explore the formation environments of non-RPE, \emph{r}-I, and \emph{r}-II stars and their distribution of [Eu/Fe] ratios in \textit{in-situ} and accreted components. Section \ref{sec:discussion} discusses how we extract the information imprinted in the MW's stars. Section \ref{sec:conclusions} provides our conclusions.

\section{Method} \label{sec:method}
\subsection{Code}

We performed a cosmological zoom-in simulation of a MW-like galaxy \citep{Hirai2022} with the $N$-body/SPH code \textsc{asura} \citep{Saitoh08, Saitoh09}. \textsc{asura} computes gravity with the tree method \citep{Barnes86} and hydrodynamics with the density-independent SPH method \citep{Saitoh13}. We adopted cooling and heating functions from 10 to 10$^9$\,K computed with  \textsc{cloudy} version 13.05 \citep{Ferland13}. Ultraviolet-background heating \citep{Haardt12} and self-shielding \citep{Rahmati13} were also implemented in this simulation. Once a gas particle satisfies the conditions of density ($>\,100$\,cm$^{-3}$), temperature ($<\,1000$\,K), and a converging flow ($\nabla\cdot\bm{v}\,<\,0$), it stochastically forms a simple stellar population star particle with the initial mass function of \citet{Chabrier03} from 0.1 to 100 $M_{\sun}$ \citep[e.g.,][]{Okamoto03, Hirai2021}.

We implemented stellar feedback to heat the interstellar medium (ISM). Stars more massive than 8 $M_{\sun}$ heat the surrounding gas within the Str\"{o}mgren radius \citep{Stromgren1939} to 10$^4$ K \citep{Fujii21}. When a star particle reaches the end of its lifetime\footnote{{The stellar lifetime was computed using the lifetime table given in \citet{Portinari1998}.}}, it distributes the energy with 10$^{51}$ erg to surrounding gas particles with the mechanical feedback model incorporating both momentum and thermal energy \citep{Hopkins18}. Elements are distributed by core-collapse supernovae (CCSNe) \citep{nomoto13}, type Ia supernovae (SNe Ia) \citep{Seitenzahl13}, asymptotic giant branch (AGB) stars \citep{Karakas10}, and NSMs {\citep{Wanajo14,Fujibayashi2023}}. We assumed delay-time distributions of SNe Ia \citep{Maoz2012} and NSMs \citep{Dominik2012} with a power law index of $-$1, with a minimum delay time ($t_{\rm{min}}$) of 40 Myr and 10 Myr, respectively. We assumed that 0.2\% of stars with 8 to 20 $M_{\sun}$ produce NSMs, and fixed the yield to reproduce the observed mean of [Eu/Fe] \citep[= +0.4;][]{Holmbeck2020}. The estimated Eu yield is $10^{-4}\,M_{\sun}$. These rates and Eu yields are within the uncertainties of the rates and yields of NSMs from observations \citep[e.g.,][]{Abbott2017ApJ, Kawaguchi2018}. {This approach has been adopted in several previous studies \citep[e.g.,][]{Ishimaru2015, vandeVoort2015, Hirai2015}.} These models were compiled in the Chemical Evolution Library \textsc{celib} \citep{Saitoh17}. Once the elements were distributed by each event, they were diffused into the ISM \citep{Hirai2017b}.

\subsection{Initial Conditions}\label{sec:ic}

We first selected a MW-like halo from a dark matter-only cosmological simulation with a box size of (36 Mpc)$^3$ using Amiga's Halo Finder \citep[\textsc{ahf},][]{Gill04, Knollmann09}. The 
cosmological parameters we adopted are: $\Omega_{\rm{m}}\,=\,0.308$, $\Omega_{\Lambda}\,=\,0.692$, $\Omega_{\rm{b}}\,=\,0.0484$, and $H_{0}\,=\,67.8$ km$\>$s$^{-1}$$\>$Mpc$^{-1}$ \citep{Planck16}. The mass of the selected halo was $1.2\times10^{12}\,M_{\sun}$. {We selected a halo that assembled over half of the total final mass before the redshift ($z$) of 2 to compute the Milky Way-like halo motivated by the quiescent merger history of the Milky Way \citep{Ruchti2015} and other cosmological zoom-in simulations of Milky Way-like galaxies \citep[e.g.,][]{Wetzel2016, Agertz2021}. Other criteria are described in \citet{Hirai2022}.}

We then performed a high-resolution cosmological zoom-in simulation. We set dark matter and gas-particle masses to be $7.2\times10^{4}\,M_{\sun}$ and $1.3\times10^{4}\,M_{\sun}$, respectively. Gravitational softening lengths are set to 85 pc for the dark matter and 82 pc for the gas. To compare with observations, we restrict our consideration to the star particles located between 3 to 20 kpc from the galactic center to be consistent with the region where Eu abundances are typically observed \citep[e.g.,][]{Gudin2021,Shank2023}.

\section{Results} \label{sec:results}

Here we analyze the timing and formation environments of RPE and non-RPE stars, extending the analysis of \citet{Hirai2022}. These authors analyzed simulated \emph{r}-II stars within 5 kpc from the Sun, as defined in their study, and showed that over 90\% of VMP \emph{r}-II stars were formed in low-mass accreted components. They found that the gas-phase [Eu/Fe] ratios in accreted components with gas mass less than 10$^7M_{\sun}$ were greatly enhanced by an NSM. In such environments, groups of \emph{r}-II stars could be formed. Here, we include stars in larger regions, between 3 and 20 kpc from the galactic center, and add simulated \emph{r}-I and non-RPE stars to our analysis. This region has 114,773 non-RPE, 869,477 \emph{r}-I, and 3357 \emph{r}-II stars. Section \ref{sec:bp} describes these stars' formation times and birthplaces.

\subsection{Properties of the Simulated Galaxy} \label{subsec:properties}

\begin{figure}[ht!]
\epsscale{1.2}
\plotone{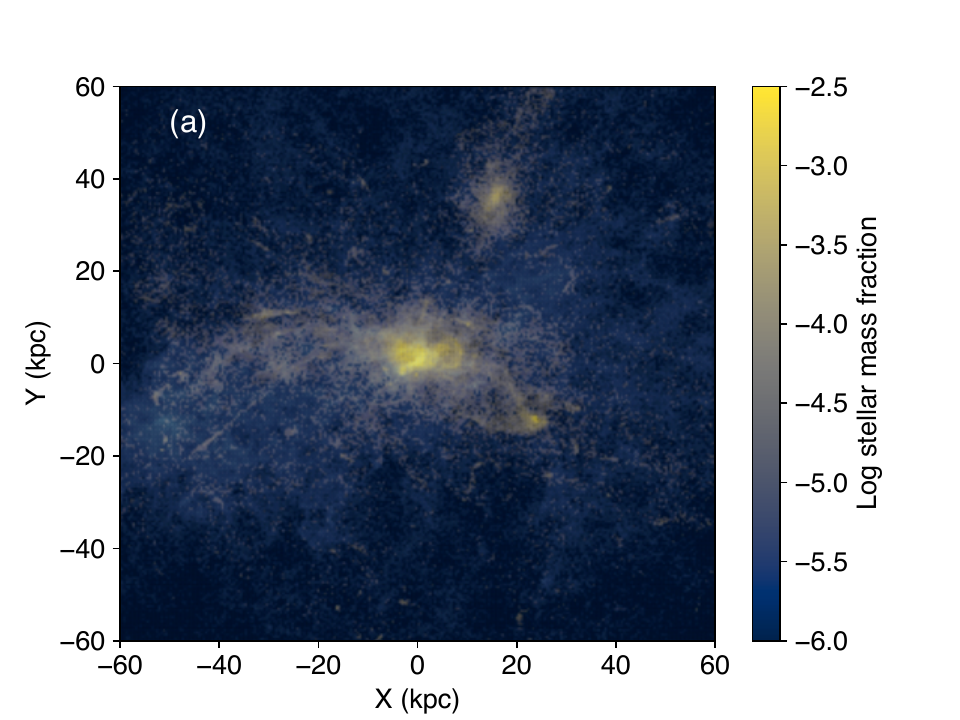}
\plotone{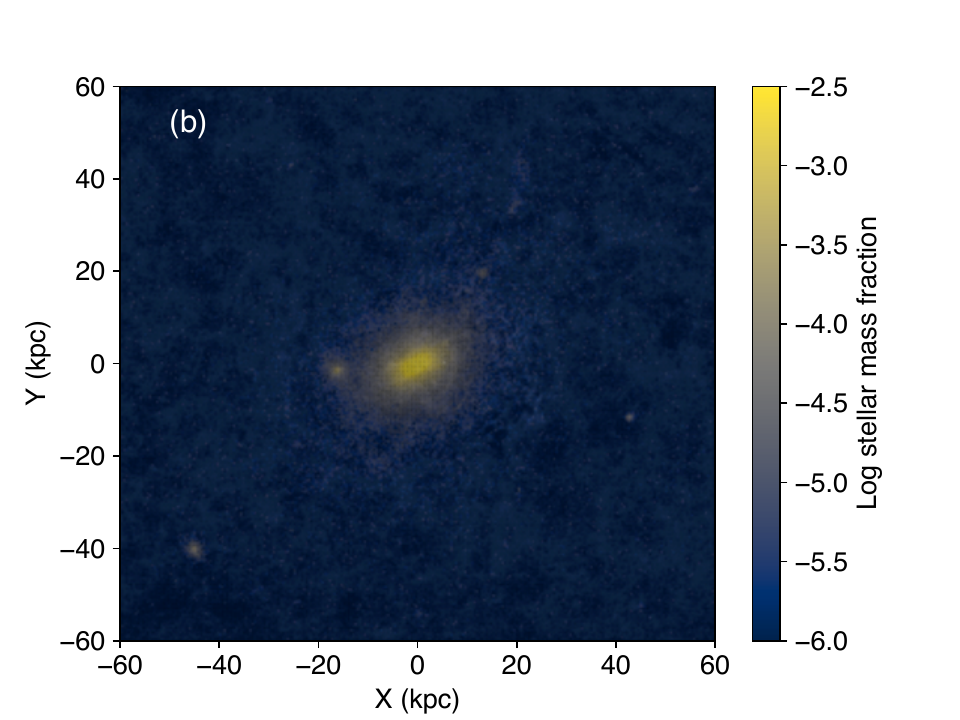}
\plotone{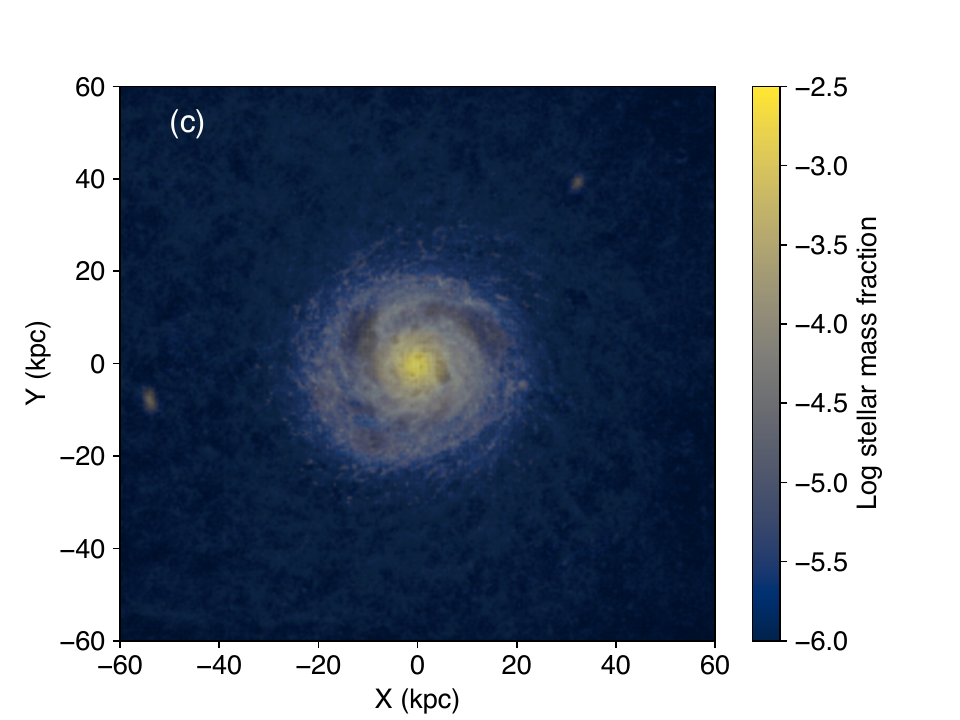}
\caption{Stellar and gas distribution of the simulated galaxy at (a) $z$ = 5.7 ($t$ = 1.0 Gyr), (b) $z$ = 1.2 ($t$ = 5.0 Gyr), and (c) $z$ = 0 ($t$ = 13.8 Gyr). The color scale, from blue to yellow, represents the stellar mass fraction with a log scale from $10^{-6.0}$ to $10^{-2.5}$.
\label{fig:gasstars}}
\end{figure}

Figure \ref{fig:gasstars} shows the stellar and gas distribution of the simulated galaxy in different epochs. Figure \ref{fig:gasstars}(a) depicts the stellar and gas distribution when most \emph{r}-II stars are formed. During this phase, the galaxy experiences frequent mergers, at $z = 5.7$ (cosmic time $t = 1.0$ Gyr). Figure \ref{fig:gasstars}(b) shows a snapshot when the star formation in the simulated galaxy is temporarily quenched, owing to gas expulsion by SNe feedback, at $z = 1.2$ (cosmic time $t = 5.0$ Gyr). After this phase, the simulated galaxy does not experience major mergers, and the cold gas's continuous accretion forms disk stars. Figure \ref{fig:gasstars}(c) is a snapshot at $z$ = 0 (cosmic time $t = 13.8$ Gyr). As shown in this figure, a spiral galaxy has been formed. In addition, there are several satellite galaxies around the central galaxy \citep{Hirai2024}; detailed interpretations of their properties await higher-resolution simulations. The stellar metallicity distribution and [$\alpha$/Fe] ratios in this simulation are shown in \citet{Hirai2022}.

\begin{figure}[ht!]
\epsscale{1.2}
\plotone{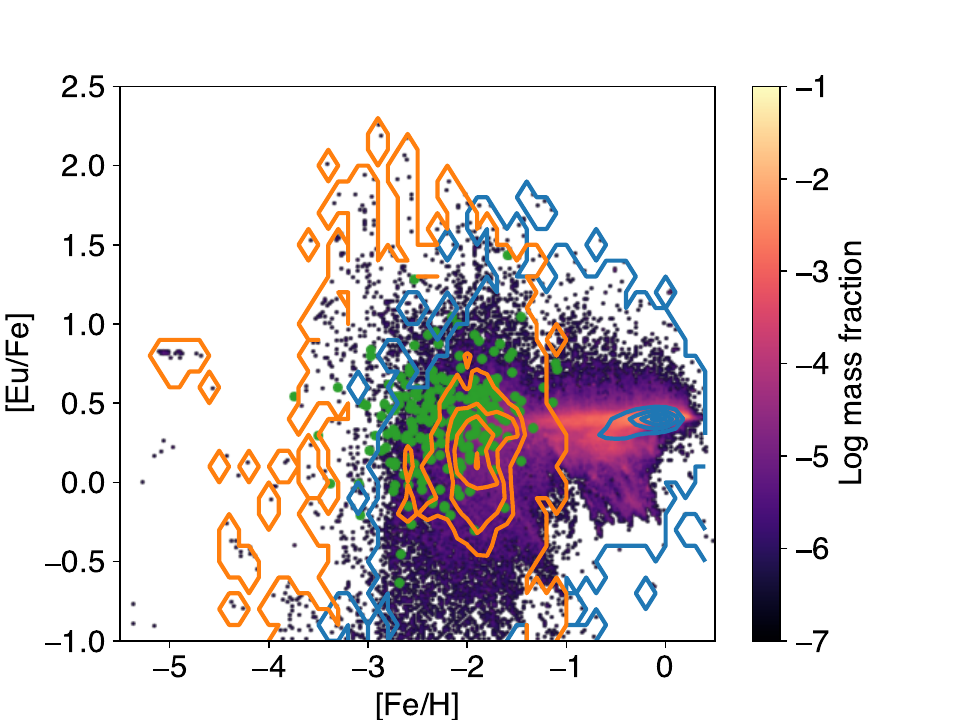}
\caption{[Eu/Fe] ratios, as a function of [Fe/H], for simulated stars located between 3 to 20 kpc from the simulated galactic center. The color scale, from {black} to yellow, represents the stellar mass fraction with a log scale from $10^{-6}$ to $10^{-2}$ in each bin of width 0.02 $\times$ 0.01\,dex$^{2}$. {Blue and orange contours represent the number density of \textit{in-situ} and accreted stars in each bin of width 0.1 $\times$ 0.1\,dex$^{2}$. Some extremely metal-poor stars are categorized as neither accreted nor \textit{in-situ} because the birthplace halo was not detected by the halo finder. Green plots are the observed RPA data \citep{Hansen2018, Sakari2018, Ezzeddine2020, Holmbeck2020}.}
\label{fig:eufe}}
\end{figure}

Simulated VMP stars exhibit star-to-star scatter of their [Eu/Fe] ratios, as shown in Figure \ref{fig:eufe}. We classify stars by the values of their [Eu/Fe] ratios, following the definitions used by the RPA. For the VMP stars with [Fe/H] $\leq -2$, \citet{Hirai2022} found that \emph{r}-II stars are preferentially formed in accreted sub-galactic fragments that are enhanced in \emph{r}-process elements at $z\gtrsim1.6$. Because of the rarity of NSMs, the gas is highly inhomogeneous at this phase, and the stellar [Eu/Fe] abundances vary among stars. Indeed, stars with [Eu/Fe] $<\,-$1 are also formed in this simulation. However, such stars are generally undetectable in observational datasets because their Eu lines are too weak for any meaningful [Eu/Fe] estimates or upper limits. 

For stars with [Fe/H] $\gtrsim-2$, the [Eu/Fe] ratios exhibit less scatter, and a relatively constant mean [Eu/Fe] ratio as a function of [Fe/H]. \citet{Hirai2022} found that most of the \emph{r}-II stars {with [Fe/H] $\gtrsim-2$} are formed in locally \emph{r}-process-enhanced gas in the main halo. The flat trend of the simulated [Eu/Fe] ratios for [Fe/H] $\gtrsim-2$ is due to the assumed delay-time distributions of NSMs and SNe Ia, which are almost identical in this simulation. In contrast, observed [Eu/Fe] ratios show a decreasing trend toward higher metallicity \citep[e.g.,][]{Guiglion2018}.  This difference can be resolved if the delay times of SNe Ia are $\sim$1 Gyr longer than those of NSMs \citep{Wanajo2021}. {The clump of simulated stars with $-$0.3 $<$ [Eu/Fe] $<$ $+$0.3 in $-$0.50 $<$ [Fe/H] $<$ $+$0.25 are formed \textit{in situ} in a pocket of gas heavily polluted by SNe Ia.}

\begin{figure}[ht!]
\epsscale{1.2}
\plotone{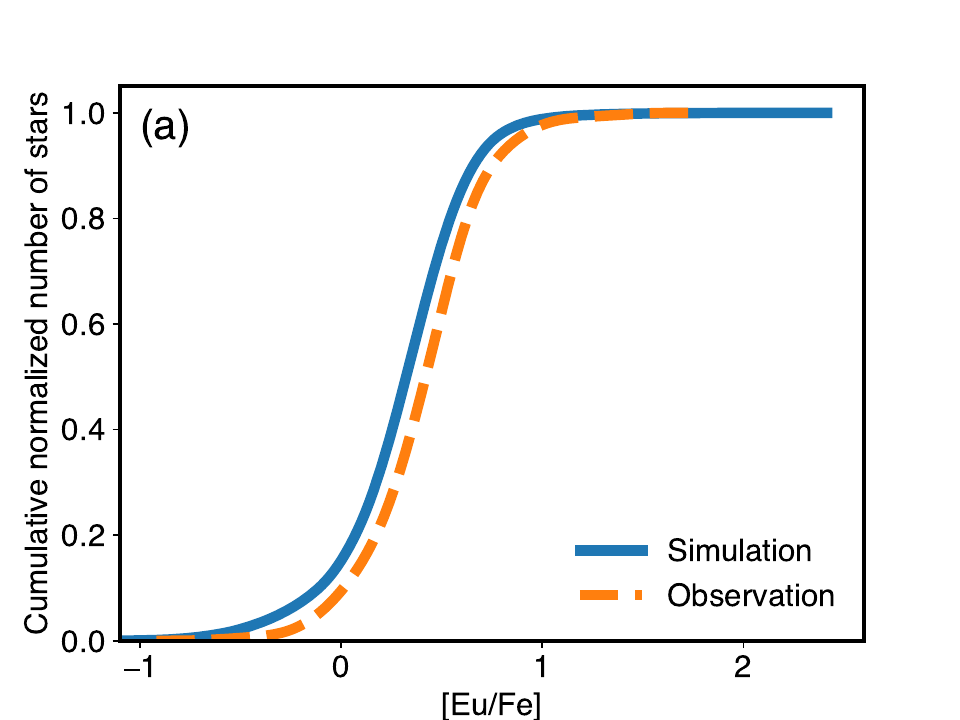}
\plotone{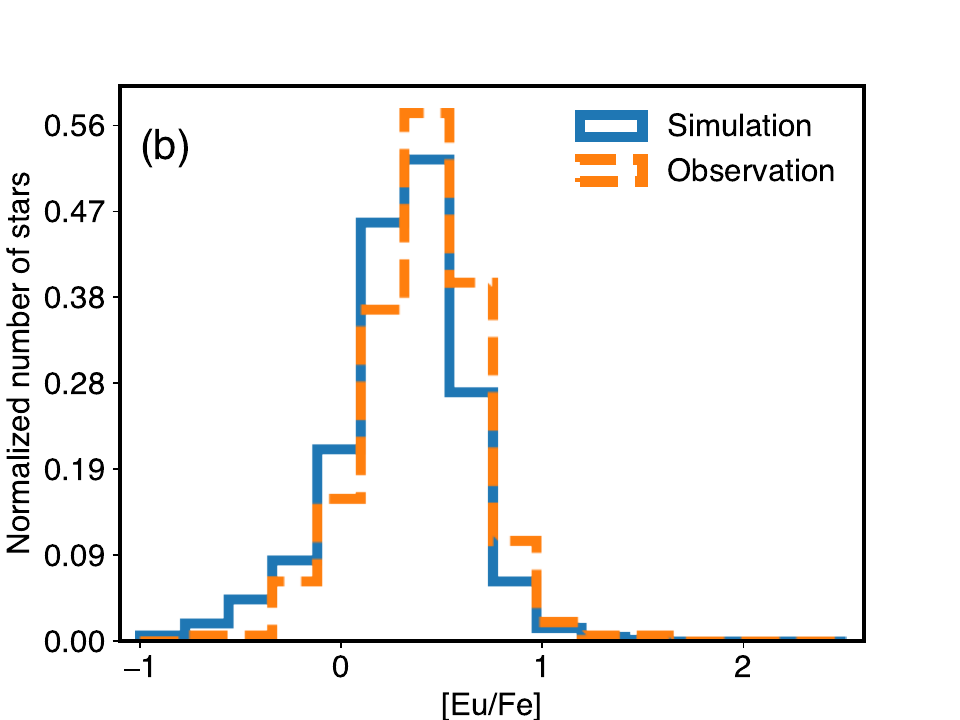}
\caption{(a) Cumulative distribution and (b) histogram of [Eu/Fe] ratios in the simulated (solid-blue line) and observed (dashed-orange line) stars. The simulation data are taken from 3 $<\,r$ (kpc) $<$ 20, $-3.5<$ [Fe/H] $<-1$, and [Eu/Fe] $>-0.8$. Typical observational errors of $\sigma_{\rm{[Eu/Fe]}}$ = 0.20\,dex are added to the simulation data. The observed data are taken from \citet{Hansen2018, Sakari2018, Ezzeddine2020, Holmbeck2020}.
\label{fig:eufe_dist}}
\end{figure}

The simulated [Eu/Fe] distribution is very similar to the observed one, as shown in Figures \ref{fig:eufe_dist}(a) and (b).  To compare with observations, we chose stars with $-3.5<$ [Fe/H] $<-1$ and [Eu/Fe] $>-0.8$ following the observed [Fe/H] and [Eu/Fe] range given in \citet{Holmbeck2020}. We have added a typical uncertainty ($\sigma_{\rm{[Eu/Fe]}}$) of 0.20\,dex to the [Eu/Fe] abundance ratios of the simulated data, as reported by \citet{Holmbeck2020} based on RPA observations, to compare the simulated [Eu/Fe] with these observations. The number fractions of simulated (observed) \emph{r}-II, \emph{r}-I, and non-RPE stars are 0.07 (0.10), 0.46 (0.56), and 0.47 (0.33), respectively. The standard deviations of simulated (observed) [Eu/Fe] distribution are 0.32 (0.29)\,dex. {We describe the effects of the mass resolution on the number fractions in Appendix \ref{app:A}.}

\subsection{Formation Times and Birthplaces of Stars}\label{sec:bp}
\begin{figure}[ht!]
\epsscale{1.2}
\plotone{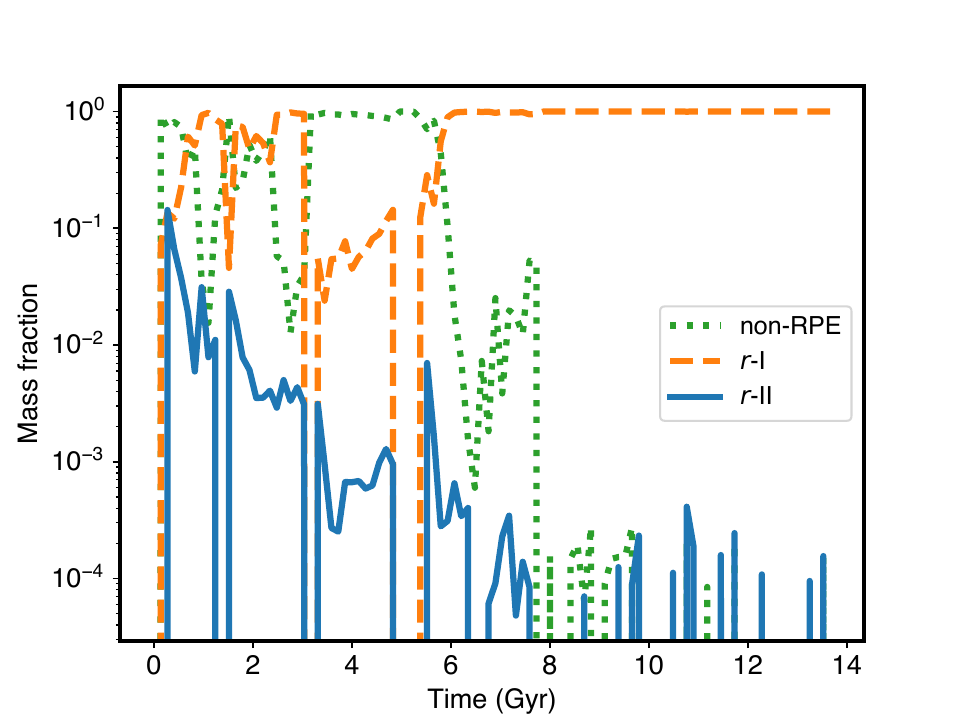}
\caption{Mass fraction of non-RPE (dotted-green line), \emph{r}-I (dashed-orange line), and \emph{r}-II (solid-blue line) stars, as a function of formation time.
\label{fig:fvstime}}
\end{figure}

\begin{figure}[ht!]
\epsscale{1.2}
\plotone{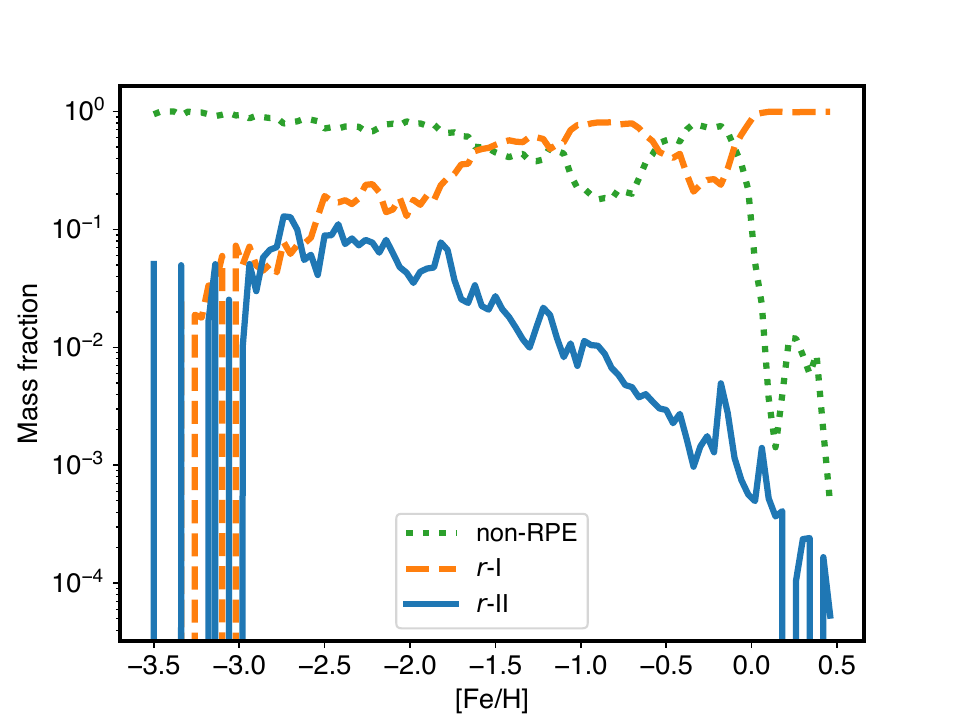}
\caption{Same as Figure \ref{fig:fvstime}, but for the mass fraction as a function of [Fe/H].
\label{fig:fvsfeh}}
\end{figure}

\begin{figure}[ht!]
\epsscale{1.2}
\plotone{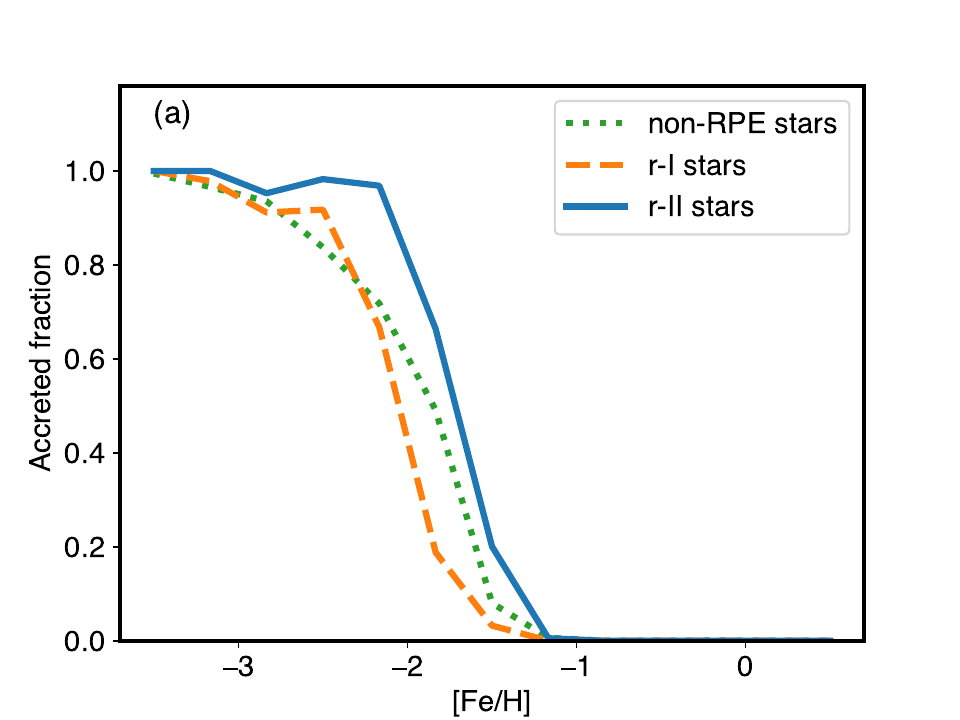}
\plotone{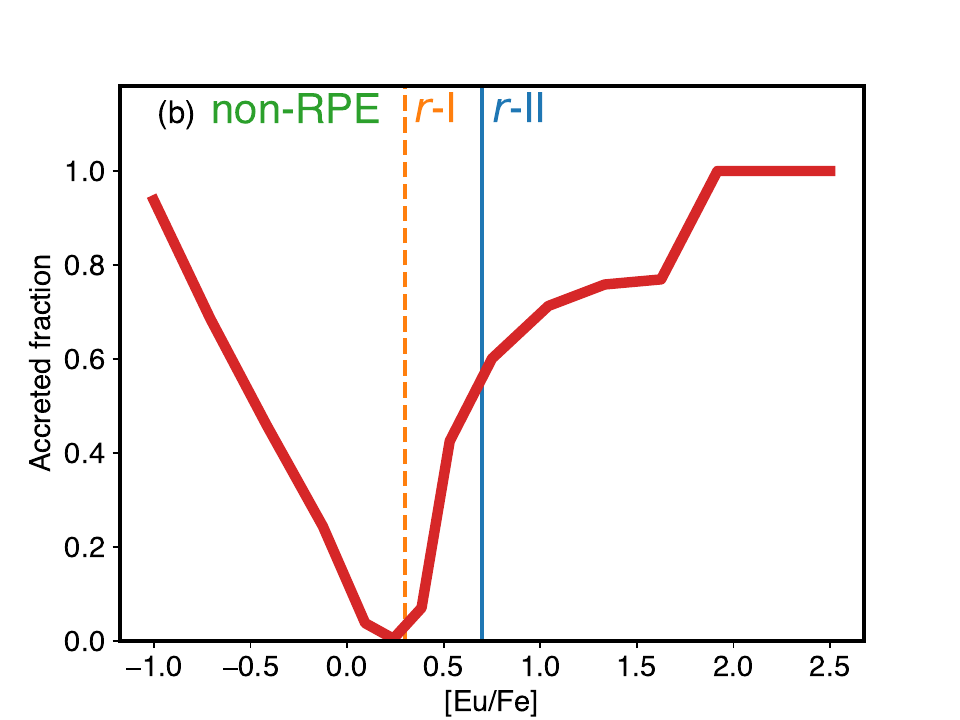}
\caption{Mass fractions of stars formed in accreted components, as a function of (a) [Fe/H], and (b) [Eu/Fe] (solid-red line). In panel (a), the dotted-green, dashed-orange, and solid-blue lines represent non-RPE, \emph{r}-I, and \emph{r}-II stars, respectively. Dashed-orange and solid-blue lines {in panel (b)} depict [Eu/Fe] = $+$0.3 and $+$0.7, respectively, which divide the non-RPE, \emph{r}-I, and \emph{r}-II classifications. 
\label{fig:facc}}
\end{figure}

\begin{figure}[ht!]
\epsscale{1.2}	
 \plotone{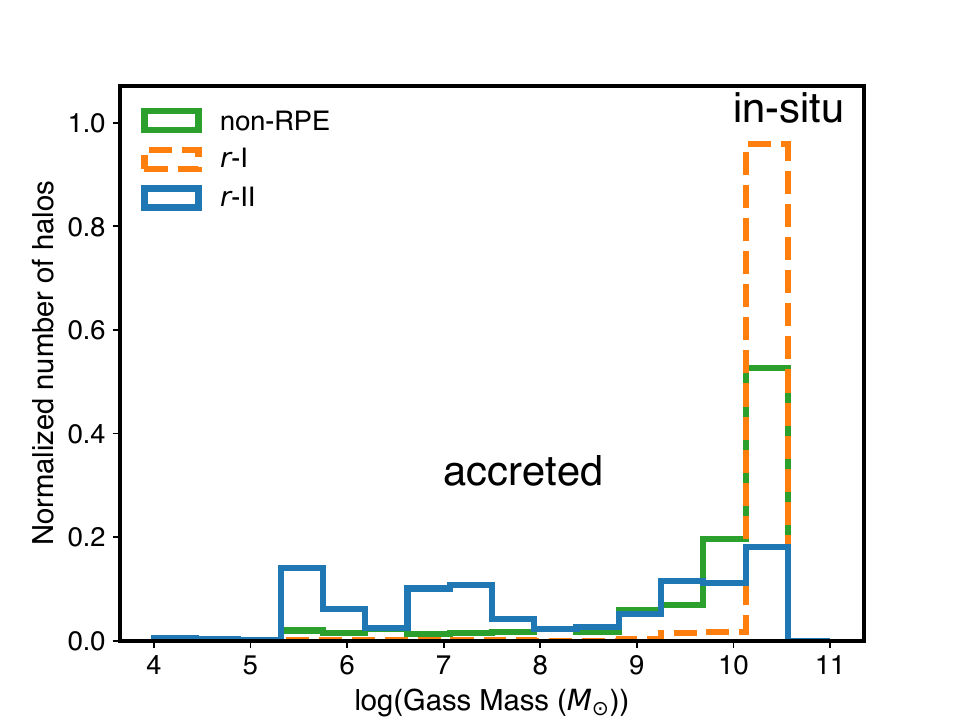}
\caption{Gas-mass distribution of halos at the cosmic time when each star formed. Green, orange, and blue histograms represent non-RPE, \emph{r}-I, and \emph{r}-II stars, respectively.}
    \label{fig:bpmass}
\end{figure}

Here we find a clear difference in the formation times and birthplaces between non-RPE/\emph{r}-II and \emph{r}-I stars (see Figures \ref{fig:fvstime}--\ref{fig:bpmass}). This difference comes from the formation environments of these stars. Because of the rarity of NSMs, most low-mass sub-halos do not host NSMs. These sub-halos have very low [Eu/Fe] (subsolar) ratios and form non-RPE stars. However, once an NSM occurs in such sub-halos, [Eu/Fe] ratios are rapidly enhanced and form \emph{r}-II stars. In contrast, \emph{r}-I stars are mainly formed in a well-mixed ISM. Below, we detail the formation times and birthplaces of non-RPE, \emph{r}-I, and \emph{r}-II stars.

Most \emph{r}-II and non-RPE stars are formed within 6 Gyr from the beginning of the simulation, while \emph{r}-I stars are formed throughout the evolutionary history of the simulated galaxy. Figure \ref{fig:fvstime} shows the mass fraction of the non-RPE, \emph{r}-I, and \emph{r}-II stars as a function of the formation time. From inspection, non-RPE or \emph{r}-I stars are the dominant contributors to the mass fraction for cosmic time $t\,\leq\, $6 Gyr. In the later phases ($t\,>\, $6 Gyr), almost all of the RPE stars are formed as \emph{r}-I stars. The mass fraction of \emph{r}-II stars is always $\lesssim 10$\% than that of non-RPE or \emph{r}-I stars. Notably, few \emph{r}-II stars are formed for $t\,>$ 8 Gyr. The mean cosmic ages are 3.5 Gyr for non-RPE stars, 8.0 Gyr for \emph{r}-I stars, and 1.8 Gyr for \emph{r}-II stars. {We describe the possible effects of the mass assembly history in Appendix \ref{app:B}.}

Figure \ref{fig:fvsfeh} shows the mass fraction of non-RPE, \emph{r}-I, and \emph{r}-II stars as a function of [Fe/H]. For [Fe/H] $< -1.54$, non-RPE stars dominate the mass fraction due to the delayed production of Eu by NSMs. Non-RPE stars with [Fe/H] $>-1$ are formed by the contribution of SNe Ia. The fraction of \emph{r}-I stars increases as a function of [Fe/H], following the increasing number of NSMs. The fraction of \emph{r}-II stars peaks at [Fe/H] $=-2.7$. Around this metallicity, most \emph{r}-II stars are formed in dwarf galaxies with gas mass $\lesssim10^7M_{\sun}$. 

Figure \ref{fig:facc}(a) shows the fraction of stars formed in accreted components (accreted fraction) as a function of [Fe/H]. The accreted fraction decreases toward higher metallicity, as reported in \citet{Hirai2022}. The percentages of stars formed in accreted components are 97\% (non-RPE), 98\% (\emph{r}-I), and 100\% (\emph{r}-II) at [Fe/H] = $-$3. These values decrease to 78\% (non-RPE), 67\% (\emph{r}-I), and 97\% (\emph{r}-II) at [Fe/H] = $-$2. Almost all stars are formed \textit{in-situ} for [Fe/H] $>-$1. Notably, the \emph{r}-II stars have higher accreted fractions compared to other stars with [Fe/H]$\,<-\,$1. The accreted fraction of non-RPE stars is also larger than that of \emph{r}-I stars for $-2\,<$ [Fe/H] $<\,-1$. 

RPE and non-RPE stars exhibit apparent differences in the trend of accreted fraction as a function of [Eu/Fe]. Figure \ref{fig:facc}(b) depicts the accreted fraction as a function of [Eu/Fe]. For non-RPE stars, the accreted fraction decreases as a function of [Eu/Fe] and reaches a minimum at [Eu/Fe] = $+$0.24. In contrast, RPE stars exhibit an increasing trend in the accreted fraction as a function of [Eu/Fe]. This result means that non-RPE stars with lower [Eu/Fe] and RPE stars with higher [Eu/Fe] tend to be formed in accreted satellite galaxies with their gas depleted or enhanced in [Eu/Fe], {respectively,} rather than being due to local inhomogeneity in the \textit{in-situ} component.

The mass distribution of gas contained in individual halos at the cosmic time when each star is formed depends on the degree of \emph{r}-process enhancement, as shown in Figure \ref{fig:bpmass}. Halos with $M_{\rm{gas}}\,>\,10^{10}\,M_{\sun}$ correspond to the \textit{in-situ} halo in most cases. We find that 96.6\% of \emph{r}-I stars are formed in the halo with $10^{10}\,M_{\sun}$, while the percentages for non-RPE stars and \emph{r}-II stars are 57.2\% and 21.4\%, respectively. This difference is mainly caused by the difference in the formation times of these stars. After 6 Gyr from the beginning of the simulation, most stars are formed as \emph{r}-I stars (Figure \ref{fig:fvstime}). Star formation in this phase mainly occurs in the simulated galactic disk. Because the disk is formed within the main halo, the total gas mass of all \emph{r}-I stars' birthplaces is estimated to be high. Hence, the available Eu enrichment is spread across a larger mass, making it more difficult for r-II stars to form. Instead, the \emph{r}-II stars tend to form in less massive halos with higher Eu concentrations.
The number fraction of stars formed in halos with $M_{\rm{gas}}\,<\,10^8\,M_{\sun}$ is 49.0\% for \emph{r}-II stars, while those for \emph{r}-I and non-RPE stars are 0.5\% and 10.8\%, respectively. These results imply that \emph{r}-II stars are preferentially formed in low gas-mass halos, while most \emph{r}-I stars tend to be formed in the \textit{in-situ}, main halo.

These findings imply that there should be large numbers of r-I stars present in the galactic disk, in particular the metal-poor portion of the thick disk, in the metal-weak thick disk/Atari disk component \citep{Norris1985, Carollo2007, Beers2014, Mardini2022}, and principally in any larger dwarf galaxy. In fact, stars with moderate [Eu/Fe] ratios have already been found in Gaia-Sausage-Enceladus \citep{Ou2024}, and several r-I stars have been identified in the Sagittarius dwarf galaxy {\citep{Ou2025}}.

\subsection{[Eu/Fe] Ratios in the \textit{In-situ} and Accreted Components}\label{sec:acc}

\begin{figure}[ht!]
\epsscale{1.2}
\plotone{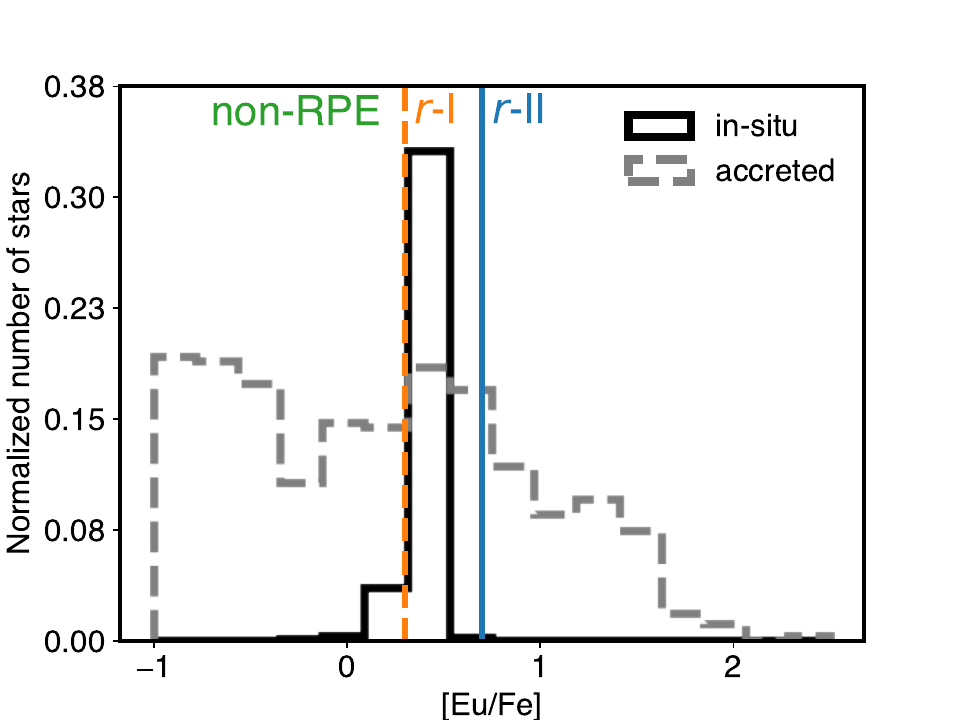}
\caption{The [Eu/Fe] distribution in the \textit{in-situ} (solid-black line) and accreted components (dashed-gray line).  Dashed-orange and solid-blue lines depict [Eu/Fe] = $+$0.3 and $+$0.7, respectively, which divide the non-RPE, \emph{r}-I, and \emph{r}-II classifications.
\label{fig:eufe_insitu}}
\end{figure}

The \textit{in-situ} and accreted components exhibit significantly different [Eu/Fe] distributions. Figure \ref{fig:eufe_insitu} shows the [Eu/Fe] distribution in the \textit{in-situ} and accreted components. The \textit{in-situ} components do not have significant dispersion ($\sigma_{\rm{[Eu/Fe]}}$ = 0.09\,dex), while the accreted components have a large dispersion in their [Eu/Fe] ratios ($\sigma_{\rm{[Eu/Fe]}}$ = 0.43\,dex). This result indicates that the star-to-star scatter of [Eu/Fe] in the MW arises from variations in the [Eu/Fe] ratios of the accreted components.

This difference is also reflected in the number fraction of \emph{r}-II and non-RPE stars. Table \ref{tab:list} shows the number fractions of \emph{r}-II, \emph{r}-I, and non-RPE stars in the \textit{in-situ} and accreted components in different mass ranges. Note that $M_{*}$ is the peak stellar mass of the accreted component, whose metallicity is lower than that inferred from the dwarf galaxy mass-metallicity relation \citep{Kirby2013}. In the \textit{in-situ} component, only 0.2\% of the stars are classified as \emph{r}-II, while more than 4\% of the stars in accreted components are \emph{r}-II. Non-RPE stars in the \textit{in-situ} component are also much lower (9.7\%) than in the accreted components ($>$ 67.2\%).

\begin{deluxetable}{cccccc}[htbp]
\tablecaption{Number Fractions of \emph{r}-II, \emph{r}-I, and Non-RPE Stars in the \textit{in-situ} and Accreted Components\label{tab:list}}
\tablewidth{0pt}
\tablehead{
    \colhead{Origin} & \colhead{$M_*$} & \colhead{$\langle {\rm [Fe/H]} \rangle$} & \colhead{$f_{r\rm{II}}$ }&\colhead{$f_{r\rm{I}}$ }&\colhead{$f_{\rm{nonRPE}}$ }\\
    \colhead{}&\colhead{($M_{\sun}$)}&\colhead{}&\colhead{}&\colhead{}&\colhead{}}
\decimals
\startdata		
            \textit{In-situ}& $>10^9$ & $-0.26$ & 0.002 & 0.901 & 0.097\\
		  Accreted & $>10^7$ & $-1.95$  & 0.073 & 0.247 & 0.680\\
            Accreted & $10^6$--$10^7$ & $-2.42$  & 0.096 & 0.232 & 0.672\\
            Accreted & $10^5$--$10^6$ & $-3.48$  & 0.042 & 0.028 & 0.930\\
\enddata
\tablecomments{From left to right, the columns show the origin of the components, the ranges of the peak stellar mass, the median [Fe/H] ($\langle {\rm [Fe/H]} \rangle$), fractions of \emph{r}-II ($f_{r\rm{II}}$), \emph{r}-I ($f_{r\rm{I}}$), and non-RPE ($f_{\rm{nonRPE}}$) stars, respectively.}
\end{deluxetable}

\begin{figure*}[ht!]
\epsscale{1.1}
    \plottwo{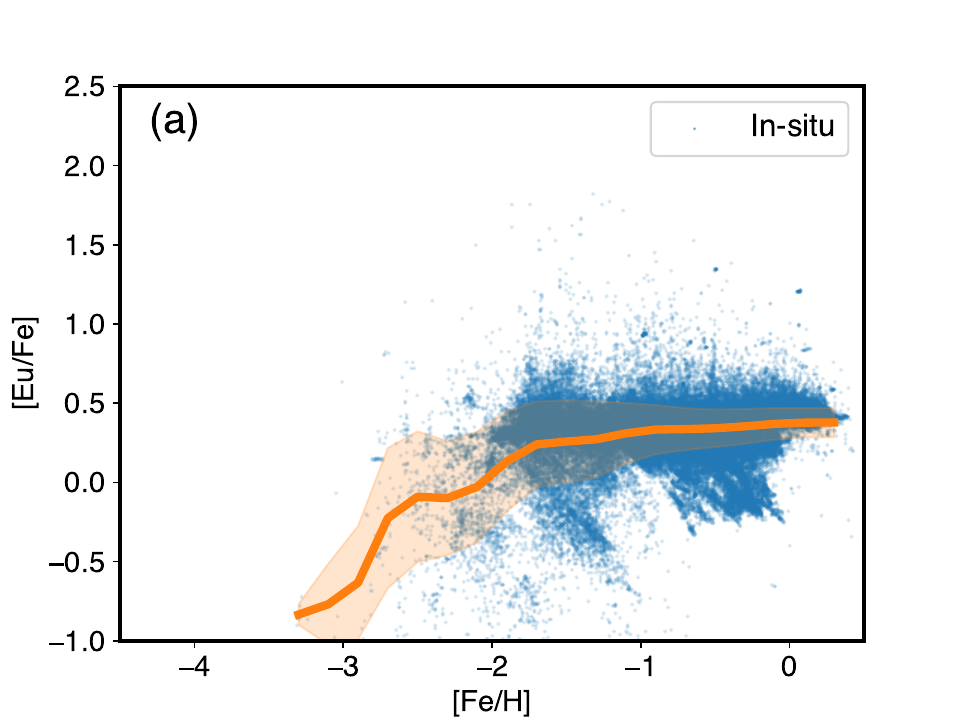}{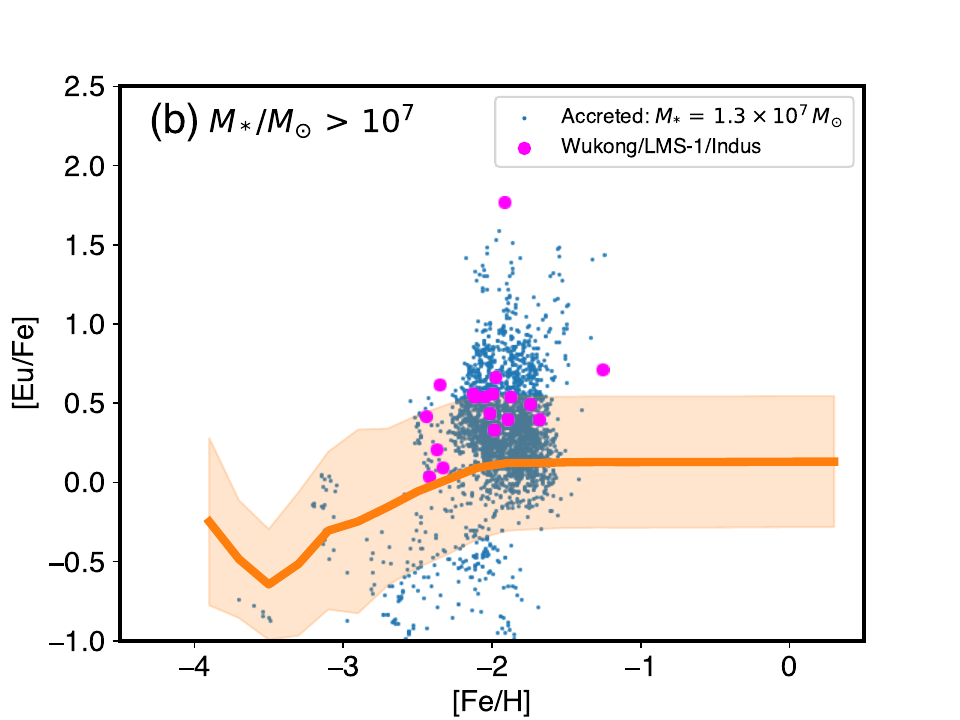}
    \plottwo{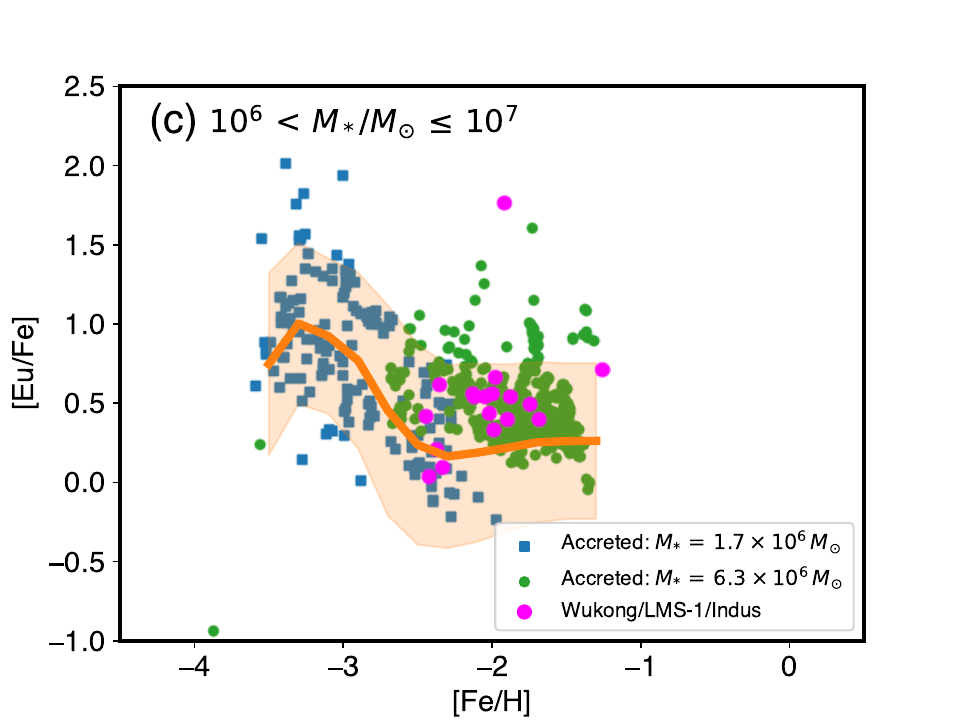}{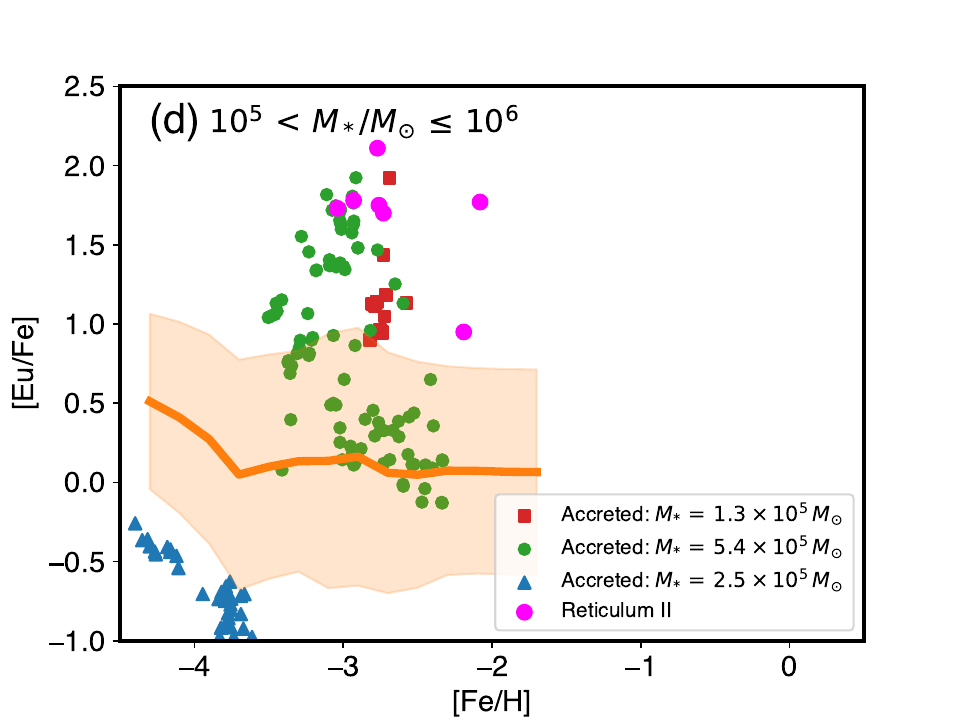}
    \caption{[Eu/Fe] ratios, as a function of [Fe/H], for the \textit{in-situ} and accreted components. The panels represent (a) the \textit{in-situ} component, the accreted components with (b) $M_*/M_{\sun}\,>\,10^7$ (accreted halo with $M_*=1.3\times10^7\,M_{\sun}$), those with (c) $10^6\,<M_*/M_{\sun}\leq\,10^7$ (accreted halo with $M_*=1.7\times10^6\,M_{\sun}$: blue squares, and with $M_*=6.3\times10^6\,M_{\sun}$: green circles), and those with (d) $10^5\,<M_*/M_{\sun}\leq\,10^6$ (accreted halo with $M_*=1.3\times10^5\,M_{\sun}$: red squares, accreted halo with $M_*=5.4\times10^5\,M_{\sun}$: green circles,
    accreted halo with $M_*=2.5\times10^5\,M_{\sun}$: blue triangles). The orange-solid lines and shaded area represent the mean [Eu/Fe] and standard deviation of [Eu/Fe] as a function of [Fe/H] for stars with [Eu/Fe] $>-1$ in all components in each mass range. The magenta plots show (b, c) Wukong/LMS-1 \citep{Limberg2024}, Indus \citep{Ji2020, Hansen2021}, and (d) Reticulum II \citep{Ji2016, Roederer2016}.}
    \label{fig:accreted}
\end{figure*}

The [Eu/Fe] ratios in each accreted dwarf galaxy exhibit characteristic behavior depending on their mass. Figure \ref{fig:accreted} shows [Eu/Fe] as a function of [Fe/H] in each of these accreted systems. We find that the region with 3 $<r/(\rm{kpc})<$ 20 comprises 168 accreted components with $M_{*}>10^5\,M_{\sun}$. 

Figure \ref{fig:accreted}(a) shows [Eu/Fe] ratios in stars formed \textit{in-situ}. This panel shows that most stars have [Fe/H] $>$ $-$3; no \emph{r}-II stars are found below this metallicity in our simulation. Because the \textit{in-situ} component is relatively massive compared to the accreted components, its chemical evolution in terms of [Fe/H] is faster than the other components. Therefore, Eu appears at a relatively higher metallicity. Also, the disk has formed in the \textit{in-situ} halo. Most stars with [Fe/H] $>\,-$1 are in the disk.

Figure \ref{fig:accreted}(b) shows [Eu/Fe] ratios in the accreted components with
$M_{*}$ $>$ 10$^7\,M_{\sun}$, of which 13 are the most massive accreted components. As shown in this figure, most stars are in the metallicity range $-$3 $\lesssim$ [Fe/H] $\lesssim -$1. Also, there is a star-to-star scatter in the [Eu/Fe] ratios. Because these accreted components are the most massive among the accreted ones, the chemical evolution of these systems is faster than for the lower-mass systems. The metallicity is already at [Fe/H] $\approx -$3 when the first NSM occurs in most of these systems. Thus, only three \emph{r}-II stars are below [Fe/H] = $-$3. After these systems are accreted to the most massive main halo, star formation is quenched, suppressing the formation of stars with [Fe/H] $> -$1. Since these systems are sufficiently massive to cause local inhomogeneities in the Eu distribution, all systems exhibit star-to-star scatter in their [Eu/Fe] ratios.

Most accreted components with $10^6\,<M_*/M_{\sun}\leq\,10^7$ exhibit star-to-star scatter in their [Eu/Fe] ratios for $-4 \lesssim$ [Fe/H] $\lesssim -1$. In this mass range, we identify 24 components. Figure \ref{fig:accreted}(c) shows two examples of accreted components in this mass range. As shown in this figure, both halos have dispersions of [Eu/Fe] ratios of over one dex. However, \emph{r}-II stars appear in different metallicity ranges in these halos. This difference arises from the differences in the times when the first stars with Eu are formed. The first RPE stars are formed at 0.4 Gyr in the halo with 1.7$\times10^6 M_{\sun}$ (blue squares). At that time, the [Fe/H] of this halo is between $-$3.6 and $-$2.8. On the other hand, the halo with 6.3$\times10^6 M_{\sun}$ (green circles) forms its first RPE stars at 0.7 Gyr. Because of the pre-enriched ISM around this halo, the metallicity has already reached [Fe/H] $\approx-2.5$ at 0.7 Gyr.

Figure \ref{fig:accreted}(d) shows three examples of [Eu/Fe] vs. [Fe/H] in the accreted components with $10^5\,<M_*/M_{\sun}\leq\,10^6$. This mass range is populated by 132 halos. We divide these mass ranges into three categories, based on their [Eu/Fe] distributions: (i) enhanced (all the stars are \emph{r}-II), (ii) scattered (there are both RPE and non-RPE stars in a group), and (iii) very low (all the stars are non-RPE). Among these 132 halos, two halos are enhanced, and eight halos are scattered. The remaining halos also display very low [Eu/Fe] ratios. Still, among these halos, 13 halos have stars with [Eu/Fe] $> -1.0$.

These differences arise from the individual star-formation histories and the formation times of NSMs. Since these halos have peak stellar masses of 10$^{5}$--10$^{6}$ $M_{\sun}$, and are accreted to the simulated MW within 1.2 Gyr, the expected number of NSMs is unity or less. This explains why 92\% of the halos in this mass range display low abundances of \emph{r}-process elements. If an NSM occurs in a halo, and it still has enough gas to form stars, RPE stars are formed. Halos categorized as enhanced exhibit single-star formation events after an NSM. Because of the small gas mass ($\lesssim 10^{7} M_{\sun}$) for these halos, the gas-phase [Eu/Fe] ratios are all enhanced \citep[see Figure 9(a) in][]{Hirai2022} after an NSM; thus, star-formation events that take place after an NSM forms RPE stars.

Halos with a dispersion of [Eu/Fe] over one dex have at least two episodes of star formation. In the case where an NSM occurs before the first peak of star formation, the [Eu/Fe] ratios exhibit a decreasing trend toward higher metallicity, as shown, e.g., in the halo plotted with green circles in Figure \ref{fig:accreted}(d). In contrast, halos show enhanced [Eu/Fe] at a higher metallicity if an NSM occurs between the two episodes of star formation.

\section{Discussion} \label{sec:discussion}

\subsection{Formation Scenarios of \emph{r}-II, \emph{r}-I, and Non-RPE Stars}\label{sec:form}

In Section \ref{sec:bp}, we found a clear difference in the formation mechanisms between non-RPE/\emph{r}-II and \emph{r}-I stars. Following our results on the formation of stars with \emph{r}-process elements, we now provide a physical interpretation of the classification of \emph{r}-II, \emph{r}-I, and non-RPE stars. 

From inspection of Figure \ref{fig:facc}(b), we find that stars in each category exhibit different trends in the fraction of stars formed in accreted components as a function of [Eu/Fe]. According to this Figure, \emph{r}-II stars tend to form in low-mass dwarf galaxies (Figure \ref{fig:bpmass}) that host NSMs. In these systems, a single NSM can significantly enhance the \emph{r}-process abundances. Such a signature has been observed in the UFD Reticulum\,II \citep{Ji2016}.
Non-RPE stars with lower [Eu/Fe] also tend to be formed in accreted dwarf galaxies that do not have NSMs (Figure \ref{fig:facc}(b)). As argued in Section \ref{sec:acc}, most low-mass components have low Eu abundances because no NSMs occurred in these systems. Because low-mass sub-halos have zero or one NSM, some systems are highly enhanced in Eu, while others have not experienced any such enrichment. Several studies have shown that an NSM event can enhance the Eu abundance sufficiently to form \emph{r}-II stars \citep[e.g.,][]{Ji2016,Safarzadeh2017,Tarumi2020, Jeon2021, Wanajo2021, Hirai2022}. Low levels of \emph{r}-process element abundances are found in Local Group UFDs (\citealt{Frebel2018}, and references therein).

In contrast, the accreted fraction of \emph{r}-I stars is, on average, less than 0.5 (Figure \ref{fig:facc}(b)). This result means that \emph{r}-I stars are mainly formed in a well-mixed ISM, such as in the disk, that has experienced ample \textit{in-situ} star formation. Because our \textit{in-situ} component hosts many NSMs with a longer duration of star formation than that of any of the accreted components, the [Eu/Fe] distribution in the ISM is homogeneous. This results in the much smaller scatter in [Eu/Fe] ratios that produce stars mostly within the \emph{r}-I regime (Figure \ref{fig:eufe_insitu}).

\subsection{Formation of Stars with Extreme Eu Abundances}\label{sec:ext}

Our simulation also includes stars with extreme abundances of \emph{r}-process elements. Some fall into the category of metal-poor \emph{r}-III stars with [Eu/Fe] $> +2.0$ and $-3.4<[\rm{Fe/H}]<-2.5$. Others appear to be more metal-rich \emph{r}-II stars with [Fe/H] $\sim-1.0$ \citep[see, e.g.,][]{Xie2024}. Yet others are \emph{r}-process element free, as they formed from gas that was never enriched by an \emph{r}-process event. \citet{Cain2020} first identified a VMP star (J1521-3538, [Fe/H] = $-2.80$) with [Eu/Fe] = $+2.23$. Because of its highly eccentric orbit, they argued that its origin is an accreted dwarf galaxy. Most recently, \citet{Roederer2024} have analyzed a new VMP \emph{r}-III star (J2213-5137; [Fe/H] = $-2.20$) with the highest known [Eu/Fe] among VMP stars found to date ([Eu/Fe] = $+2.45$).  

Our simulation predicts that the percentage of \emph{r}-III stars is 6.1$\times10^{-4}$\%, suggesting that these stars may well be difficult to find. The most \emph{r}-process-enhanced star in our simulation within 3 $<r/(\rm{kpc})<$ 20 displays [Eu/Fe] = $+$2.45 (at $r = 7$ kpc). Outside of this range, the highest ratio is [Eu/Fe] = $+$3.58 (at \emph{r} = 60 kpc). We find that all \emph{r}-III stars are from accreted components. Among them, two r-III stars are from a massive accreted component with $M_{*}$ = 1.3$\times$10$^7$ $M_{\sun}$, while the other four stars are from low-mass accreted components with $M_{*}$ $< 10^5 M_{\sun}$. This result suggests that accreted dwarf galaxies are the exclusive sites for forming \emph{r}-III stars. Most of them are formed in the low-mass accreted components that are highly enhanced in \emph{r}-process elements, while local inhomogeneities in massive accreted components still allows for the formation of \emph{r}-III stars.

As shown in Figure \ref{fig:fvsfeh}, we find that a few metal-rich ([Fe/H] $>-$1) \emph{r}-II stars formed in the \textit{in-situ} galactic disk, but these were rare occurrences.  
From inspection of Figure \ref{fig:facc}(a), all \emph{r}-II stars in this simulation are formed \textit{in-situ} {for this metallicity range}. These stars are formed in the locally Eu-enhanced gas clump around an NSM occurring in the galactic disk. However, only 0.08\% of the stars are classified as metal-rich \emph{r}-II stars in our simulation. Regarding observations, \citet{Xie2024} found a \emph{r}-II star (J0206+4941) at [Fe/H] = $-0.54$ in the thin disk and formed \textit{in-situ}, based on its kinematic signature. The discovery of J0206+4941 has allowed calculating the percentage of metal-rich \emph{r}-II stars in the MW to be 0.03\%, i.e., one per 2664 stars in the literature \citep[SAGA database,][]{Suda2008, Suda2011, Yamada2013}. \citet{Hirai2017b} measured the timescale to dilute and mix the local \emph{r}-process enhancement near an NSM, and found that it was erased within 40 Myr on average \citep[also see Figure 11 of ][]{Hirai2022}. These results mean that the short timescale of metal dilution suppresses the formation of metal-rich \emph{r}-II stars.

In contrast to the most extreme RPE stars, our simulation also has some stars without Eu. We find that 0.0058\% of the stars are Eu free. However, as a result of a general level of metal mixing, most stars will form with at least some Eu. This tiny percentage of Eu-free stars suggests that stars without neutron-capture elements may never be found. 

\subsection{Diagnostics to Estimate Stellar Mass of Accreted Components\label{sec:eumg}}
Europium abundances may be useful to distinguish the stellar masses of accreted components. In Section \ref{sec:acc}, we have shown that the metallicity range, trends in [Eu/Fe], and fractions of \emph{r}-II stars differ among accreted components with different masses. For the largest accreted components with $>10^7\,M_{\sun}$, most stars have metallicities in the range $-$3 $<$ [Fe/H] $<$ $-$1, with scatter in [Eu/Fe] (Figure \ref{fig:accreted}(b)). On the other hand, \emph{r}-II stars with [Fe/H] $<-3$ are found in accreted components with $\leq10^7\,M_{\sun}$ (Figures \ref{fig:accreted}(c) and (d)). Because of their low gas masses, a single NSM event enhances the entire sub-halo, resulting in downward trends of [Eu/Fe] toward higher metallicity.

As argued in Section \ref{sec:acc}, a single NSM event can enhance gas phase [Eu/Fe] ratios in the accreted components of our simulation because of their low gas mass. Indeed, enhancement of \emph{r}-process elements in extremely metal-poor stars has been confirmed in the Reticulum II UFD \citep{Ji2016, Ji2023, Roederer2016}. Figure \ref{fig:accreted}(d) compares [Eu/Fe] as a function of [Fe/H] in the Reticulum II UFD and that in accreted components with $M_{*}< 10^6 M_{\sun}$. From inspection, our accreted components with $M_*=1.3\times10^5\,M_{\sun}$ (red squares) and $M_*=5.4\times10^5\,M_{\sun}$\footnote{Note that $M_*$ represents the peak stellar mass.} (green circles) exhibit similar levels of enhancements of [Eu/Fe] ratios
to those observed in Reticulum II. A similar conclusion can be reached for Tucana III \citep{Hansen2017}. 
In addition, this agreement further suggests that CDTGs found amongst \emph{r}-II stars \citep{Roederer2018,Gudin2021,Shank2023, Hattori2023} could also originate from similar low-mass accreted early dwarf galaxies.

\begin{figure}[ht!]
\epsscale{1.2}	
 \plotone{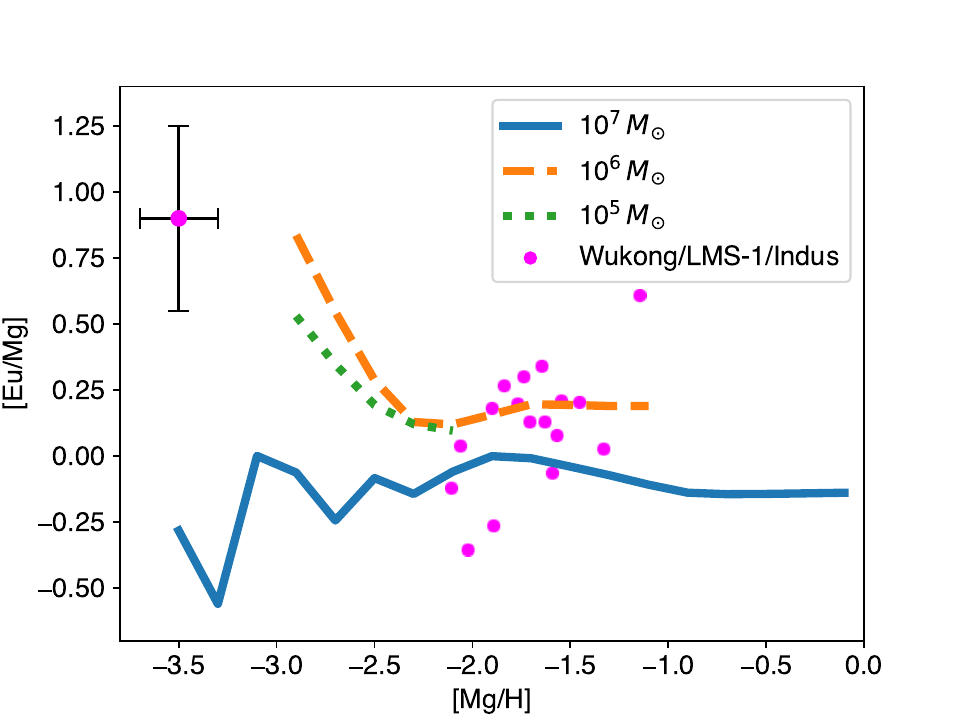}
\caption{Mean [Eu/Mg], as a function of [Mg/H], for accreted components with peak stellar masses of $M_*/M_{\sun}\,>\,10^7$ (blue-solid line), $10^6\,<M_*/M_{\sun}\leq\,10^7$ (orange-dashed line), and $10^5\,<M_*/M_{\sun}\leq\,10^6$ (green-dotted line), respectively. The mean [Eu/Mg] is computed for stars with [Eu/Mg] $>-1$. The magenta plots represent Wukong/LMS-1 \citep{Limberg2024} and Indus \citep{Ji2020}. {Typical measurement errors \citep{Limberg2024} are shown in the top left corner.}}
    \label{fig:eumg}
\end{figure}

In the following, we attempt to calculate the stellar mass of the accreted components, based on our derived chemical abundances. In general, the [Eu/Mg] ratio as a function of [Mg/H] clearly distinguishes the stellar masses of accreted components. Since CCSNe are the main source of Mg, we can eliminate the uncertainties of delay-time distributions of SNe Ia in [Eu/Mg] versus [Mg/H]. Figure \ref{fig:eumg} shows the mean [Eu/Mg] as a function of [Mg/H] for stars with [Eu/Mg] $> -1$ {to compare with observations. It is difficult to measure the abundances of stars with [Eu/Mg] $< -1$ in the current observations.} From inspection, the mean [Eu/Mg] ratios exhibit downward trends toward higher [Mg/H] in the range $-3<$ [Mg/H] $< -2$ for low-mass accreted components with mass $\leq 10^7 M_{\sun}$. This trend arises from the enhancement of \emph{r}-process elements by an NSM in low gas-mass galaxies. In contrast, accreted components with stellar mass $>10^7 M_{\sun}$ show a flat to slightly increasing trend toward higher [Mg/H], due to the lack of \emph{r}-II stars with [Fe/H] $< -3$. These results suggest that [Eu/Mg] ratios can be an indicator to distinguish the low-mass ($\leq 10^7 M_{\sun}$) and massive ($>10^7 M_{\sun}$) accreted components.

Here, we propose criteria to estimate the stellar masses of accreted components: (1) \emph{r}-II stars with [Fe/H] $< -3$ are highly likely to come from low-mass accreted components with $M_{*}<10^7M_{\sun}$ (Figures \ref{fig:accreted}(c) and (d)). (2) A group of stars with a downward trend of [Eu/Mg] in $-3<$[Mg/H]$< -2$ toward higher [Mg/H] would also come from low-mass components with $M_{*}<10^7M_{\sun}$ (Figure \ref{fig:eumg}). (3) A group of stars showing an upward trend of [Eu/Fe] and [Eu/Mg] in $-3<$[Fe/H]$< -2$ toward higher metallicity possibly comes from accreted components with $M_{*}>10^7M_{\sun}$ (Figures \ref{fig:accreted}(b) and \ref{fig:eumg}). Table \ref{tab:diagnostics} summarizes these criteria.
\begin{deluxetable}{cccccc}[htpb]
\tablecaption{Diagnostics to Estimate the Stellar Mass of Accreted Components \label{tab:diagnostics}}
\tablewidth{0pt}
\tablehead{
    \colhead{Criteria} & \colhead{[Mg/H]}& \colhead{[Fe/H]}&\colhead{[Eu/Mg] }&\colhead{[Eu/Fe]} & \colhead{$M_*$}\\
    \colhead{}&\colhead{}&\colhead{}&\colhead{}&\colhead{}&\colhead{($M_{\sun}$)}}
\decimals
\startdata		
    (1) &--& $<-3$ & -- & $>+0.7$& $<10^7$\\
    (2) &$-3$ to $-2$& -- & down & --& $<10^7$\\
    (3) &$-3$ to $-2$& $-3$ to $-2$ & up & up& $>10^7$\\
\enddata
\tablecomments{From left to right, the columns show the criteria, [Mg/H], [Fe/H], [Eu/Mg], [Eu/Fe], and the estimated stellar mass ($M_*$). The ``down" and ``up" notations in columns of [Eu/Mg] and [Eu/Fe] mean downwards and upward trends toward higher metallicity, respectively.}
\end{deluxetable}

\begin{figure}[ht!]
\epsscale{1.2}	
\plotone{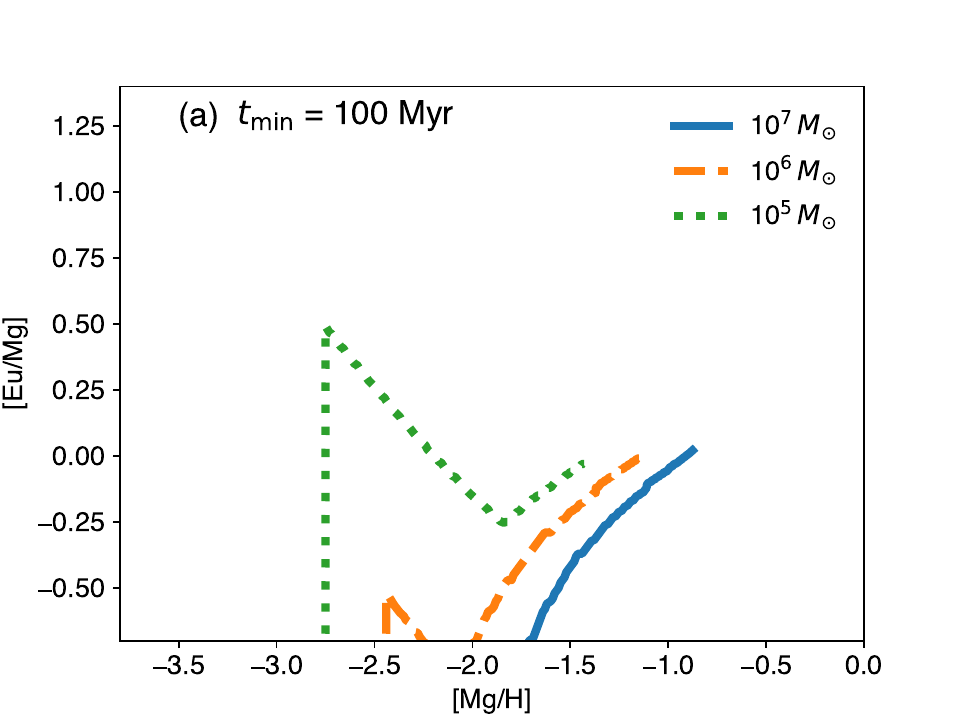}
\plotone{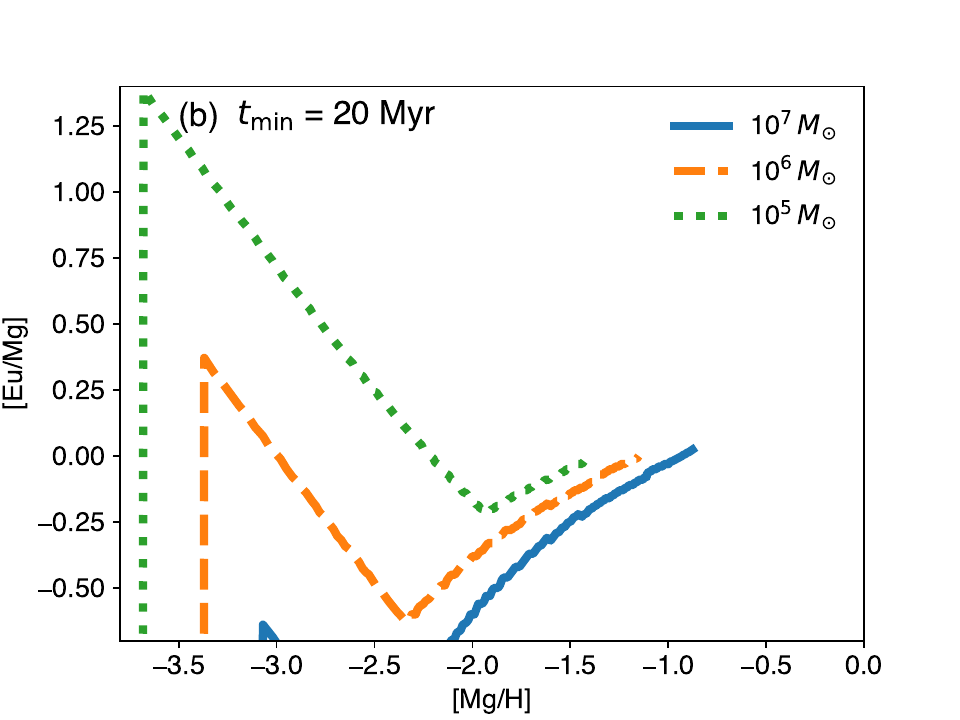}
\plotone{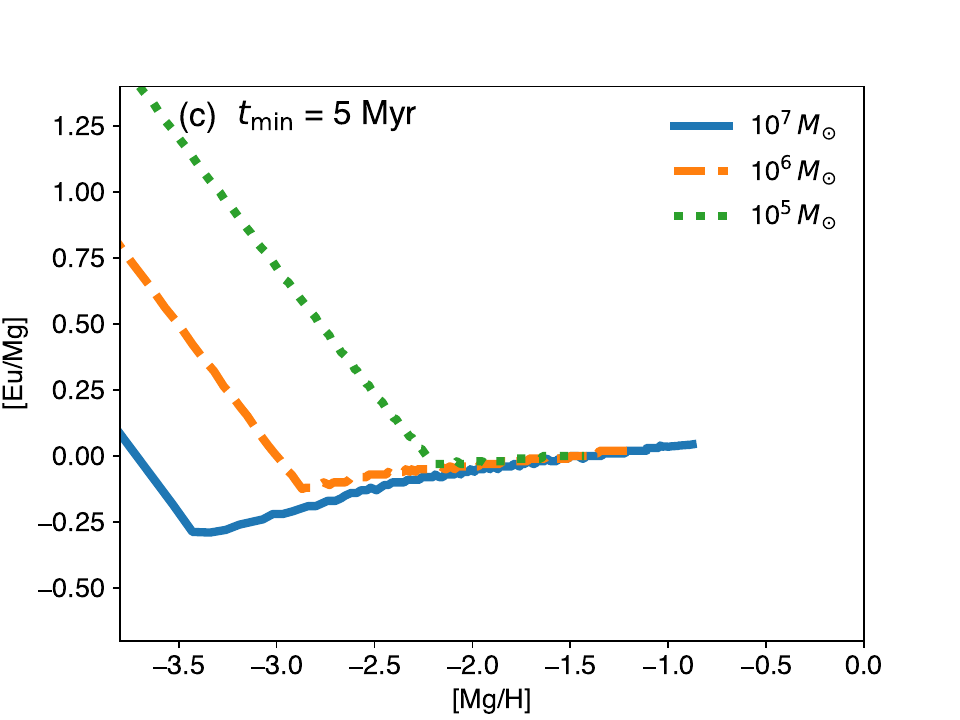}
\caption{[Eu/Mg], as a function of [Mg/H], for (a) case 1B (NSMs, $t_{\rm{min}}$ = 100 Myr), (b) case 1 (NSMs, $t_{\rm{min}}$ = 20 Myr), and (c) case 3 (CCSNe, $t_{\rm{min}}$ = 5 Myr) in \citet{Wanajo2021}. Blue-solid, orange-dashed, and green-dotted lines represent models with the final stellar mass of 10$^7$, 10$^6$, and 10$^5$ $M_{\sun}$, respectively. Each model is computed for 2 Gyr.}
    \label{fig:wanajo21}
\end{figure}

The downward trends of [Eu/Mg] ratios in low-mass accreted components are independent of the delay times of \emph{r}-process enrichment. Figure \ref{fig:wanajo21} shows one-zone chemical-evolution models assuming different delay times of \emph{r}-process enrichment: (a) case 1B (NSMs with $t_{\rm{min}}$ = 100 Myr), (b) case 1 (NSMs with $t_{\rm{min}}$ = 20 Myr),  and (c) case 3 (CCSNe) \citep[see tables 1 and A1 in ][]{Wanajo2021}. In these plots, we divide $N_{\rm{Eu}}/N_{\rm{Mg}}$\footnote{$N_{\rm{Eu}}$ and $N_{\rm{Mg}}$ are the number densities of Eu and Mg, respectively.} by the average number of NSMs for each galaxy if it is less than unity. As shown in this figure, the downward trend of [Eu/Mg] can be seen regardless of the delay times of \emph{r}-process enrichment.

The upward trend of [Eu/Mg] in massive accreted components suggests a delay of the \emph{r}-process enrichment compared to Mg. Models assuming delay times of NSMs show upward trends of [Eu/Mg] toward higher [Mg/H] (Figures \ref{fig:wanajo21}(a) and (b)). In contrast, case 3 in \citet{Wanajo2021}, assuming $t_{\rm{min}} = 5$ Myr, exhibits an almost constant [Eu/Mg] as a function of [Mg/H] (Figure \ref{fig:wanajo21}(c)). {This case corresponds to the model assuming collapsars or /magneto-rotational driven supernovae \citep[e.g.,][]{Molero2023}.} The upward trend of our simulation (the blue line in Figure \ref{fig:eumg}) is unclear because we plot stars with [Eu/Mg] $>-1$. Including stars with [Eu/Mg] $\leq-1$ clarifies the upward trend.

Upward trends of [Eu/Mg] as a function of metallicity have been reported in several dwarf galaxies. \citet{Limberg2024} found that [Eu/Mg] ratios in the stellar stream LMS-1/Wukong \citep{Yuan2020, Naidu2020} continuously increase toward higher metallicity (magenta plots in Figures \ref{fig:accreted}(b), (c) and \ref{fig:eumg}). \citet{Ou2024} reported that an accreted component Gaia-Sausage-Enceladus has a clear rise of [Eu/Mg] as a function of [Mg/H]. In the Large Magellanic Cloud, \citet{Chiti2024} showed that stars with [Fe/H] $< -2.5$ did not exhibit \emph{r}-process enhancements, while stars with $-2.5 <$ [Fe/H] $< -1.5$ were enhanced in their [Eu/Fe] ratios \citep{Reggiani2021}. Simulated accreted components with $M_{*} > 10^7 M_{\sun}$ also lack \emph{r}-II stars with [Fe/H] $< -2.5$ (Figure \ref{fig:accreted}(b)). This property results in the upward trend of [Eu/Fe] and [Eu/Mg] toward higher metallicity. These results suggest that upward trends of [Eu/Fe] and [Eu/Mg] could be an indicator to detect accreted components with $M_{*} > 10^7 M_{\sun}$.

\section{Conclusions}\label{sec:conclusions}
In this study, we have performed a high-resolution cosmological zoom-in simulation of a Milky Way-like galaxy using the \textsc{asura} code. We analyze the formation environments of \emph{r}-II, \emph{r}-I, and non-RPE stars within 3 to 20 kpc from the galactic center, which mimics the region of the known \emph{r}-process metal-poor stars identified by the RPA and other searches. We find that our simulated environment exhibits a [Eu/Fe] distribution similar to that of the RPA sample to date (Figure \ref{fig:eufe_dist}). We also show the [Eu/Fe] and [Eu/Mg] ratios for \textit{in-situ} and accreted components with different mass ranges.

We find that non-RPE stars and \emph{r}-II stars are formed on similar time scales and in similar environments. Most non-RPE and \emph{r}-II stars are formed within 6 Gyr from the Big Bang (Figure \ref{fig:fvstime}), preferentially in low-gas mass accreted components (Figures \ref{fig:facc}(b) and \ref{fig:bpmass}). On the other hand, \emph{r}-I stars are continuously formed for 13 Gyr, mainly in the \textit{in-situ} component. These results indicate a clear difference in the formation environment between non-RPE/\emph{r}-II and \emph{r}-I stars. In low-mass sub-halos, an NSM can enhance the entire galaxy, forming \emph{r}-II stars. Given their small mass and the rarity of NSMs, some sub-halos will never host NSMs. In these cases, non-RPE stars are formed. For the \textit{in-situ} halo, most gas is well-mixed. In such an environment, \emph{r}-I stars can form for extended periods of time.

Furthermore, our simulation forms stars with either strongly enhanced Eu abundances or none at all. Hence, we find that all \emph{r}-III stars are formed in accreted components; the highest [Eu/Fe] ratio in our simulation is [Eu/Fe] = $+$3.58. We also find it relatively rare for any metal-rich \emph{r}-II stars to form in local inhomogeneities in the disk. Rather, these stars require formation in locally \emph{r}-process-enhanced gas clumps near a previous NSM. 
Interestingly, there are few \emph{r}-process free stars formed in our simulation. This result suggests that star formation in pristine gas without \emph{r}-process elements is infrequent, likely due to metal mixing and dilution processes. Although stars with extreme \emph{r}-process abundances are scarce in our simulation, finding actual stars with extreme abundances would greatly help to constrain the ejecta properties of NSMs and the effects of metal dilution.

In this work, we propose diagnostics to estimate the stellar mass of accreted components. 
Low-mass accreted components with $M_{*}<10^7M_{\sun}$ appear to be hosts of 
(1) \emph{r}-II stars with [Fe/H] $< -3$ or 
(2) A group of stars with a downward trend of [Eu/Mg] vs [Mg/H] across the range of $-3<$ [Mg/H] $< -2$. Massive accreted components with $M_{*}>10^7M_{\sun}$ appear to host   
(3) A group of stars with an upward trend of [Eu/Fe] and [Eu/Mg] vs [Fe/H] across the range of $-3<$ [Fe/H] $< -2$.
A large fraction of the known \emph{r}-II stars in the halo have [Fe/H] $\sim-3$, thus suggesting that they all formed in early accreted systems which is consistent with other theoretical predictions (e.g., \citealt{Brauer2019}). Overall, this confirms that observed \emph{r}-process signatures can be employed in chemical tagging of the accretion history of the Milky Way,  ultimately leading to a reconstruction of our Galaxy's assembly history.
Hence, these predictions should be further tested and applied to the chemo-dynamics of metal-poor stars in the Milky Way.

To make progress, as stars leftover from any of the accreted components are more likely to be found in the outer halo \citep[e.g.,][]{Beers2012, Font2020, Suzuki2024}, it would be desirable to extend the RPA effort to larger distances in order to uncover fainter \emph{r}-process stars. We also expect that the Legacy Survey of Space and Time will find many new dwarf galaxies and substructures \citep{Ivezic2019}. The Subaru Prime Focus Spectrograph can perform massively multiplexed spectroscopic observations for such structures in the outer halo \citep{Takada2014, Hirai2024}. Follow-up spectroscopic observations with next-generation Extremely Large Telescopes \citep[e.g.,][]{roederer2024elt} will be able to measure \emph{r}-process abundances of these stars. Comparison with high-resolution simulations, in particular those with the ability to simulate the formation of individual stars, will further improve understanding of the enrichment of \emph{r}-process elements and the accretion history of the MW.

\begin{acknowledgments}
We thank Alexander P.\ Ji for his insightful comments. This work was supported in part by JSPS KAKENHI Grant Numbers JP22KJ0157, JP25H00664, JP25K01046, JP21H04997, JP23H00127, JP23H05432, JP23H04894, JP23H04891, JP24H00027, JP24K00669, JP22K03688, JP24K07095, MEXT as ``Program for Promoting Researches on the Supercomputer Fugaku" (Structure and Evolution of the Universe Unraveled by Fusion of Simulation and AI; Grant Number JPMXP1020230406), JICFuS, grants PHY 14-30152; Physics Frontier Center/JINA Center for the Evolution of the Elements (JINA-CEE), and OISE-1927130: The International Research Network for Nuclear Astrophysics (IReNA), awarded by the US National Science Foundation (NSF). Y.S.L.\ acknowledges support from the National Research Foundation (NRF) of Korea grant funded by the Ministry of Science and ICT (RS-2024-00333766). I.U.R.\ acknowledges support from NSF grant AST~2205847 and NASA Astrophysics Data Analysis Program grant 80NSSC21K0627. The work of V.M.P.\ is supported by NOIRLab, which is managed by the Association of Universities for Research in Astronomy (AURA) under a cooperative agreement with the U.S. NSF. T.T.H.\ acknowledges support from the Swedish Research Council (VR 2021-05556). R.E.\ acknowledges support from NSF grant AST
2206263 and NASA Astrophysics Theory Program grant
80NSSC24K089. A.F. acknowledges support from NSF-AAG grant AST-2307436. E.M.H.\ acknowledges this work performed under the auspices
of the U.S.\ Department of Energy by Lawrence Livermore National Laboratory under Contract DE-AC52-07NA27344.
This document has been approved for release under LLNL-JRNL-2000637.

Numerical computations and analysis were carried out on Cray XC50 and computers at the Center for Computational Astrophysics, National Astronomical Observatory of Japan and the  Yukawa Institute Computer Facility. This research has made use of NASA's Astrophysics Data System.
\end{acknowledgments}

\software{astropy \citep{Astropy2013,Astropy2018},
          CELib \citep{Saitoh17},
          Cloudy \citep{Ferland13}
          }
\appendix
\section{{Effects of Mass Resolution on the Fractions of \emph{r}-II, \emph{r}-I, and Non-RPE Stars}}\label{app:A}

{We computed the fractions of \emph{r}-II, \emph{r}-I, and non-RPE stars in our simulations with different resolutions. In addition to the fiducial resolution shown in the main text, we computed the fraction in a low-resolution run with a gas-particle mass of 1.1 $\times$ 10$^5\,M_{\sun}$ and the gravitational softening length of 163 pc for the same halo with our fiducial run. Table \ref{tab:resolution} shows the fractions for these stars. We computed the fractions in all zoomed-in areas of the simulations. As shown in this table, fractions of \emph{r}-II and non-RPE stars are larger in the higher resolution one. This result is mainly because low-mass halos, where most of these stars are formed, are well resolved in the high-resolution case. The formation of \emph{r}-II stars in the in-situ halo is rare regardless of the resolution. These results suggest that resolution does not affect our conclusions on the formation mechanism and diagnostics to estimate the mass of accreted components.}

\begin{deluxetable}{cccc}[htbp]
\tablecaption{{Number Fractions of \emph{r}-II, \emph{r}-I, and Non-RPE Stars in Simulations with Different Resolution}\label{tab:resolution}}
\tablewidth{0pt}
\tablehead{
    \colhead{$m_{\rm{gas}}$} & \colhead{$f_{r\rm{II}}$ }&\colhead{$f_{r\rm{I}}$ }&\colhead{$f_{\rm{nonRPE}}$ }\\
    \colhead{($M_{\sun}$)}&\colhead{}&\colhead{}&\colhead{}}
\decimals
\startdata		
            $1.3\times10^4$ & 0.008 & 0.4833 & 0.5087\\
		  $1.1\times10^5$ & 0.004 & 0.6439 & 0.3521\\
\enddata
\tablecomments{{From left to right, the gas-particle mass, fractions of \emph{r}-II ($f_{r\rm{II}}$), \emph{r}-I ($f_{r\rm{I}}$), and non-RPE ($f_{\rm{nonRPE}}$) stars, respectively.}}
\end{deluxetable}

\section{{Effects of the Mass Assembly History}}\label{app:B}
{The mass assembly history could affect the timings of the formation of RPE and non-RPE stars. If there are galaxy assembly more recently, the age distribution of \emph{r}-II stars could be younger than our simulation. The zoomed-in halo has been selected based on the mass assembly history. We selected the halo without a major merger since $z$ = 2. Thanks to this halo selection criterion, we expect that the formation timings are not largely affected by the mass assembly history.}

{The fraction of \emph{r}-II stars could indicate the assembly of low-mass halos. Since these stars are mainly formed in low-mass halos, galaxies assembled by the larger fractions of low-mass halos should have larger fractions of \emph{r}-II stars. Comparison with observations could constrain the mass assembly history. As shown in Section \ref{subsec:properties}, the simulated (observed) number fraction of \emph{r}-II stars in 3 $< r$/kpc $<$ 20 is similar: 0.07 (0.10). This result means that our simulation has a similar mass assembly history for low-mass halos to the Milky Way.}

\bibliography{references}{}
\bibliographystyle{aasjournalv7}
\end{document}